\documentclass[twocolumn,showpacs,preprintnumbers,amsmath,amssymb]{revtex4}

\usepackage{graphicx}
\usepackage{dcolumn}
\usepackage{bm}

\begin{document}


\title{Observation of the Anisotropy of 10~TeV Primary Cosmic Ray
  Nuclei Flux with the Super-Kamiokande-I Detector}

\newcounter{foots}

\newcommand{\authoraticrr}{$^{1}$}
\newcommand{\authoratrccn}{$^{2}$}
\newcommand{\authoratbu}{$^{3}$}
\newcommand{\authoratbnl}{$^{4}$}
\newcommand{\authoratuci}{$^{5}$}
\newcommand{\authoratcsu}{$^{6}$}
\newcommand{\authoratduke}{$^{7}$}
\newcommand{\authoratgmu}{$^{8}$}
\newcommand{\authoratgifu}{$^{9}$}
\newcommand{\authoratuh}{$^{10}$}
\newcommand{\authoratiu}{$^{11}$}
\newcommand{\authoratkek}{$^{12}$}
\newcommand{\authoratkobe}{$^{13}$}
\newcommand{\authoratkyoto}{$^{14}$}
\newcommand{\authoratlanl}{$^{15}$}
\newcommand{\authoratlsu}{$^{16}$}
\newcommand{\authoratumd}{$^{17}$}
\newcommand{\authoratduluth}{$^{18}$}
\newcommand{\authoratmiyagi}{$^{19}$}
\newcommand{\authoratnagoya}{$^{20}$}
\newcommand{\authoratokayama}{$^{21}$}
\newcommand{\authoratosaka}{$^{22}$}
\newcommand{\authoratseoul}{$^{23}$}
\newcommand{\authoratshinshu}{$^{24}$}
\newcommand{\authoratseika}{$^{25}$}
\newcommand{\authoratshizuoka}{$^{26}$}
\newcommand{\authoratsuny}{$^{27}$}
\newcommand{\authorattohoku}{$^{28}$}
\newcommand{\authorattokai}{$^{29}$}
\newcommand{\authorattit}{$^{30}$}
\newcommand{\authorattokyo}{$^{31}$}
\newcommand{\authoratwarsaw}{$^{32}$}
\newcommand{\authoratuw}{$^{33}$}

\newcommand{\addressoficrr}[1]{$^{1}$ #1 }
\newcommand{\addressofrccn}[1]{$^{2}$ #1 }
\newcommand{\addressofbu}[1]{$^{3}$ #1 }
\newcommand{\addressofbnl}[1]{$^{4}$ #1 }
\newcommand{\addressofuci}[1]{$^{5}$ #1 }
\newcommand{\addressofcsu}[1]{$^{6}$ #1 }
\newcommand{\addressofduke}[1]{$^{7}$ #1 }
\newcommand{\addressofgmu}[1]{$^{8}$ #1 }
\newcommand{\addressofgifu}[1]{$^{9}$ #1 }
\newcommand{\addressofuh}[1]{$^{10}$ #1 }
\newcommand{\addressofiu}[1]{$^{11}$ #1 }
\newcommand{\addressofkek}[1]{$^{12}$ #1 }
\newcommand{\addressofkobe}[1]{$^{13}$ #1 }
\newcommand{\addressofkyoto}[1]{$^{14}$ #1 }
\newcommand{\addressoflanl}[1]{$^{15}$ #1 }
\newcommand{\addressoflsu}[1]{$^{16}$ #1 }
\newcommand{\addressofumd}[1]{$^{17}$ #1 }
\newcommand{\addressofduluth}[1]{$^{18}$ #1 }
\newcommand{\addressofmiyagi}[1]{$^{19}$ #1 }
\newcommand{\addressofnagoya}[1]{$^{20}$ #1 }
\newcommand{\addressofokayama}[1]{$^{21}$ #1 }
\newcommand{\addressofosaka}[1]{$^{22}$ #1 }
\newcommand{\addressofseoul}[1]{$^{23}$ #1 }
\newcommand{\addressofshinshu}[1]{$^{24}$ #1 }
\newcommand{\addressofseika}[1]{$^{25}$ #1 }
\newcommand{\addressofshizuoka}[1]{$^{26}$ #1 }
\newcommand{\addressofsuny}[1]{$^{27}$ #1 }
\newcommand{\addressoftohoku}[1]{$^{28}$ #1 }
\newcommand{\addressoftokai}[1]{$^{29}$ #1 }
\newcommand{\addressoftit}[1]{$^{30}$ #1 }
\newcommand{\addressoftokyo}[1]{$^{31}$ #1 }
\newcommand{\addressofwarsaw}[1]{$^{32}$ #1 }
\newcommand{\addressofuw}[1]{$^{33}$ #1 }

\author{
{\large The Super-Kamiokande Collaboration} \\
\bigskip
G.~Guillian$^{10,}$\footnote{Present address: Physics Department, Queen's University, Kingston, Ontario, Canada K7L 3N6},
J.~Hosaka\authoraticrr,
K.~Ishihara\authoraticrr,
J.~Kameda\authoraticrr,
Y.~Koshio\authoraticrr,
A.~Minamino\authoraticrr,
C.~Mitsuda\authoraticrr,
M.~Miura\authoraticrr, 
S.~Moriyama\authoraticrr, 
M.~Nakahata\authoraticrr
% ...
\authoratshinshu \\
\bigskip
\smallskip
\footnotesize
\it
\addressoficrr{Kamioka Observatory, Institute for Cosmic Ray Research,
  University of Tokyo, Hida, Gifu, 506-1205, Japan}\\
\addressofrccn{Research Center for Cosmic Neutrinos, Institute for
  Cosmic Ray Research, University of Tokyo, Kashiwa,Chiba 277-8582, Japan}\\
\addressofbu{Department of Physics, Boston University, Boston, MA 02215, USA}\\
\addressofbnl{Physics Department, Brookhaven National Laboratory, Upton, NY 11973, USA}\\
\addressofuci{Department of Physics and Astronomy, University of California, Irvine, Irvine, CA 92697-4575, USA }\\
\addressofcsu{Department of Physics, California State University, Dominguez Hills, Carson, CA 90747, USA}\\
\addressofduke{Department of Physics, Duke University, Durham, NC
  27708, USA}\\
\addressofgmu{Department of Physics, George Mason University, Fairfax, VA 22030, USA }\\
\addressofgifu{Department of Physics, Gifu University, Gifu, Gifu 501-1193, Japan}\\
\addressofuh{Department of Physics and Astronomy, University of Hawaii, Honolulu, HI 96822, USA}\\
\addressofiu{Department of Physics, Indiana University, Bloomington,
  IN 47405, USA}\\
\addressofkek{High Energy Accelerator Research Organization (KEK), Tsukuba, Ibaraki 305-0801, Japan }\\
\addressofkobe{Department of Physics, Kobe University, Kobe, Hyogo 657-8501, Japan}\\
\addressofkyoto{Faculty of Science, Kyoto University, Kyoto 606-8502, Japan}\\
\addressoflanl{Physics Division, P-23, Los Alamos National Laboratory, Los Alamos, NM 87544, USA }\\
\addressoflsu{Department of Physics and Astronomy, Louisiana State University, Baton Rouge, LA 70803, USA }\\
\addressofumd{Department of Physics, University of Maryland, College Park, MD 20742, USA }\\
\addressofduluth{Department of Physics, University of Minnesota
Duluth, MN 55812-2496, USA}\\
\addressofmiyagi{Department of Physics, Miyagi University of
  Education, Sendai, Miyagi 980-0845, Japan}\\
\addressofnagoya{Solar-Terrestrial Environment Laboratory, Nagoya University, Nagoya, 464-8601, Japan}\\
\addressofokayama{Okayama University, Okayama 700-8530, Japan}\\
\addressofosaka{Department of Physics, Osaka University, Toyonaka, Osaka 560-0043, Japan}\\
\addressofseoul{Department of Physics, Seoul National University, Seoul 151-742, Korea}\\
\addressofshinshu{Shinshu University, Matsumoto 390-8621, Japan}\\
\addressofseika{International and Cultural Studies, Shizuoka Seika
  College, Yaizu, Shizuoka 425-8611, Japan}\\
\addressofshizuoka{Department of Systems Engineering, Shizuoka University, Hamamatsu, Shizuoka 432-8561, Japan}\\
\addressofsuny{Department of Physics and Astronomy, State University of New York, Stony Brook, NY 11794-3800, USA}\\
\addressoftohoku{Research Center for Neutrino Science, Tohoku University, Sendai, Miyagi 980-8578, Japan}\\
\addressoftokai{Department of Physics, Tokai University, Hiratsuka, Kanagawa 259-1292, Japan}\\
\addressoftit{Department of Physics, Tokyo Institute for Technology, Meguro, Tokyo 152-8551, Japan }\\
\addressoftokyo{The University of Tokyo, Tokyo 113-0033, Japan }\\
\addressofwarsaw{Institute of Experimental Physics, Warsaw University, 00-681 Warsaw, Poland }\\
\addressofuw{Department of Physics, University of Washington, Seattle, WA 98195-1560, USA}\\
}
\affiliation{ } 

\date{\today}

\begin{abstract}

The relative sidereal variation in the arrival direction of primary
cosmic ray nuclei of median energy 10~TeV was measured using downward,
through-going muons detected with the Super-Kamiokande-I detector.
The projection of the anisotropy map onto the right ascension axis has
a first harmonic amplitude of $(6.64 \pm 0.98~\mbox{stat.} \pm
0.55~\mbox{syst.}) \times 10^{-4}$ and a phase at maximum at
$(33.2^\circ \pm 8.2^\circ~\mbox{stat.} \pm 5.1^\circ~\mbox{syst.})$
right ascension.  A sky map in equatorial coordinates indicates an
excess region in the constellation of Taurus and a deficit region
toward Virgo.  The excess region is centered at $(\alpha_T, \,
\delta_T) = (75^\circ \pm 7^\circ, \, -5^\circ \pm 9^\circ)$ with a
half opening angle $\chi_T = (39 \pm 7)^\circ$; the excess flux is
($0.104 \pm 0.020$)\% above the isotropic expectation.  The
corresponding parameters for the deficit region are $(\alpha_V, \,
\delta_V) = (205^\circ \pm 7^\circ, \, 5^\circ \pm 10^\circ)$, $\chi_V
= (54 \pm 7)^\circ$, and $(-0.094 \pm 0.014)$\%.  The data do not allow us to rule out a pure dipole form for the anisotropy (allowed at 13\% confidence level); they are better described by the excess and deficit cones described above.  We explored the implications under the assumption that the true anisotropy is not distorted too much by the analysis filter so that it is well-described by the observed excess and deficit cones. 

\end{abstract}

\pacs{95.85.Ry, 96.50.Bh}
\maketitle

\section{Introduction}

The flux of cosmic rays with energy per nucleon in the range $10^{11}
\sim 10^{14}$~eV is known to have a sidereal anisotropy of several
times $10^{-4}$.  The anisotropy is due to a combination of effects.
Compton and Getting~\cite{CG} proposed in 1935 that the motion of the
solar system relative to the rest frame of the cosmic ray plasma
should cause an energy-independent dipole anisotropy whose maximum is
in the direction of this motion.  Solar diurnal and seasonal changes
in the atmospheric temperature can induce a sidereal variation in the
cosmic ray rate~\cite{Farley}.  The anisotropy that remains after
accounting for these effects is presumably of Galactic origin, with
possible modulations due to the heliosphere (see, for example,
\cite{NAGASHIMA_1989, NFJ}) and, at the lowest energies, solar wind
and magnetic field.

In this article, we present a report on the observation of cosmic ray
anisotropy with the Super-Kamiokande I (SK-I) detector.  SK-I can make
a unique contribution to this subject because of the large overburden
and detector size.  The overburden makes SK-I sensitive to primary
cosmic ray energies normally attainable with extensive air shower
arrays, while the large statistics and excellent muon tracking
resolution enabled the creation of a two dimensional map of the
anisotropy, which is the first-published muon-based
map~\cite{imb_note}.

\section{The Detector and the Data}
\label{sec:detector_and_data}

SK-I is a 50~kiloton underground imaging water Cherenkov detector in
Kamioka, Japan at geographical coordinates $36^\circ 25^\prime
32.6^{\prime\prime}$~N, $137^\circ 18^\prime 37.1^{\prime\prime}$~E
and an altitude of $370$~m above sea level.  The vertical overburden
is about 1000 meters, or 2700 meters water equivalent.  The detector's
design was optimized for the detection of neutrinos and nucleon decay;
the placing of the detector under large overburden to shield against
cosmic ray muons is an important part of this design.  The overburden
shields all charged cosmic ray secondaries except muons with energy
above 0.8~TeV.  The portion of the detector sensitive to muons is a
cylinder of diameter 33.8~m and height 36.2~m, giving a target area
between 1000~m$^2$ and 1200~m$^2$ depending on the zenith angle.  The
average cosmic ray muon event rate was 1.8~Hz.  More details about the
SK-I detector are reported in~\cite{SK-I_NIM}.

The data used in this analysis were collected between June 1, 1996 and
May 31, 2001.  During this period, the detector was live for 1662.0
days (91.0\% of the time) and registered $2.54 \times 10^{8}$ muon
events.  Muon track reconstruction was performed with an algorithm
developed in SK-I to examine the spatial correlation between
spallation products and parent muons in the solar neutrino
analysis~\cite{ishino}.  In order to maintain the angular resolution
within $2^\circ$, the muon tracks were required to be longer than 10~m
and downward-going; 82.6\% of the events ($2.10 \times 10^{8}$)
satisfied these requirements.  The reliability of using muon tracks
for astronomical purposes was confirmed by the observation of the
shadow of the moon and the sun~\cite{kajiyama}.

The relationship between the energy of the detected muon and the
energy of the primary cosmic ray that produced it is described by a
response function (see, for example, Ref.~\cite{murakami}).  For SK-I,
the threshold muon energy is 0.8~TeV for the thinnest part of the
overburden.  The corresponding median primary cosmic ray energy is
about ten times larger~\cite{murakami}, while the spread in the
primary cosmic ray energy is about an order of magnitude above and
below the median.  SK-I is, therefore, sensitive to primary cosmic
rays with energy in the range several~TeV to several hundred TeV.

\begin{figure}
\includegraphics[width=25em]{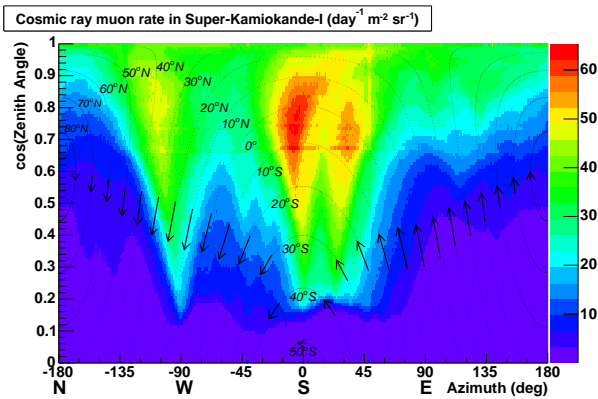}
\caption{\label{fig:overburden_norm} Event rate ($\mbox{day}^{-1}
  \mbox{m}^{-2} \mbox{s}r^{-1}$) in horizontal coordinates.  The
  dotted curves indicate contours of constant declination, while the
  arrows indicate the apparent motion of stars with the rotation of
  the earth.}
\end{figure}

The variation in the overburden along different lines of sight
explains most of the features seen in the muon event rate in the
horizontal coordinate system (Fig.~\ref{fig:overburden_norm}).  This
variation implies that the muon threshold energy, and, therefore, the
median primary cosmic ray energy, vary with direction.  A given point
in the celestial coordinate system traces out a trajectory of fixed
declination as the Earth rotates; the overburden along the line of
sight to this point varies with this motion.  For instance, the
thickness of the overburden at the apex of the declination = $0^\circ$
trajectory is about 2300 meters water equivalent, which corresponds to
a median primary cosmic ray energy of about 8~TeV.  In contrast, the
corresponding thickness and median energy for declination =
$50^\circ$~S are 3800 meters water equivalent and 20~TeV,
respectively.  The average overburden over one rotation period is,
therefore, a function of declination, which implies that the primary
cosmic ray spectrum seen by SK varies with declination.  For this reason, the
anisotropy presented here is a spectrum-weighted average over a broad
range of energies spanning several to several hundred TeV.

It is seen from Fig.~\ref{fig:overburden_norm} that the detector is
most sensitive to cosmic rays originating from the south.  During 5
years' operation of SK-I, the most sensitive direction was exposed to
every $1^\circ~\mbox{\sc ra}$ slice of the celestial sphere for an
average of about 4.6 days.  The exposure, however, was unequal for
different directions because the detector was dead for several minutes
almost every day, with occasional periods of down time lasting many
hours.  Periods of detector down time were not random, but occurred
more often during day time hours.  These periods of dead time,
accumulated over five years, introduce fluctuations in the exposure
times for different directions.  The maximum and minimum deviations
from the average exposure were $^{+1.4\%}_{-1.3\%}$.  In the celestial
coordinate analysis of Sec.~\ref{sec:celestial}, the rates were
corrected to account for different measured detector live time in
different sidereal time bins.

The atmosphere is a part of the detector in the sense that it is
responsible for converting primary cosmic rays into muons that can
penetrate the overburden.  It is a dynamic detector component because
its density changes with temperature and pressure, and the muon rate
changes accordingly.  The relative variation in the muon rate due to
atmospheric variation has relatively strong Fourier components with
frequencies corresponding to one year and one solar day.  The solar
diurnal component of the muon rate is, to some extent, modulated by a
seasonally varying signal, giving rise to spurious sidereal variation.
The magnitude of this variation is large (comparable to the true
magnitude) when the observation period is restricted seasonally.
However, averaged over exact-year periods, the spurious variation
largely cancels out; we estimate it to be only 18\% of the true
magnitude.  For this reason, we chose our data to span the exact five
year period between June 1, 1996 and June 1, 2001.  In the celestial
coordinate analysis below, we statistically subtract this spurious
variation from the observed signal using the method of Farley and
Storey~\cite{Farley}.  This subtraction introduces an uncertainty of
about 10\% to the true magnitude of the sidereal
anisotropy~\footnote{The systematic uncertainty on the true magnitude,
10\%, is significantly smaller than the magnitude of the spurious
anisotropy, 18\%, because the statistical correction technique
involves addition of phasors, which are two-dimensional vectors.
Significant cancellation occurs when the vectors are added
statistically.}.  Technical details on the subtraction of the spurious
atmospheric effect is given in Appendix~\ref{app:atmospheric_effects}.
It is also important to verify that the input to this correction is
sound; we show in Appendix~\ref{app:atmospheric} that the monthly muon
rate variation, which is primarily due to atmospheric effects, is
properly measured in Super-Kamiokande.

Another correction that can be made to the anisotropy measurement is
that for the Compton-Getting effect.  We chose not to subtract this in
our main result because the rest frame of the cosmic ray plasma -- an
important input for the subtraction -- is not known.  In principle
this can be measured with our data by measuring the seasonal change in
the anisotropy.  When the earth's orbital phase is such that the
orbital velocity moves against the bulk cosmic ray motion, the flux is
enhanced in this direction; six months later, the effect should be
smaller, as the earth moves along with the bulk motion and the
relative velocity between the observer and the cosmic ray rest frame
is at a minimum.  In practice, however, the cosmic ray rest frame was
not measurable in this manner because the seasonal change in the flux
introduced by atmospheric effects was much larger than the
Compton-Getting signal.  In the literature, two cosmic ray rest frames
are often assumed: (1) the local standard of rest (the frame of the
average motion of stars in the neighborhood of the sun); (2) the rest
frame of the local interstellar medium.  Appendix~\ref{app:cg_correct}
gives the result of anisotropy measurements with the subtraction of
the Compton-Getting effect assuming these two rest frames, and the
resulting anisotropy parameters are summarized in
tables~\ref{tab:cone_params_cg_correct} and
\ref{tab:amp_phase_corrected}.

\section{Celestial Coordinate Analysis}
\label{sec:celestial}

In this section, we describe the celestial coordinate analysis of the anisotropy.  First, the general mathematical framework of the analysis is given.  An important part of this discussion is the distortion introduced to the anisotropy map by the analysis method.  In the second subsection, the implementation of the method to the data is described.  In the third subsection, the anisotropy in one dimension (i.e. as a function of right ascension) is presented.  This is followed by the anisotropy map in two dimensions. 

\subsection{Analysis Method: Mathematical Framework and the Distortion Introduced to the Anisotropy Map} 

The number of downward-going muon events from the celestial
coordinates $(\alpha, \delta)$ observed in SK-I may be expressed as
follows:

\begin{equation}
\label{eqn:n_alpha_delta}
N(\alpha_i, \delta_j) = \sum_{k}^{N_{sid}} \left[ 1 +
  \epsilon(\alpha_i, \delta_j) \right] \cdot R(\alpha_i - \tau_k,
  \delta_j) \cdot T_k \cdot d\Omega_{i,j}
\end{equation}

\noindent The indexes $i$, $j$, and $k$ are for the right ascension,
declination, and sidereal time, $N_{sid}$ is the number of sidereal
time bins, and $T_k$ is the total detector live time in the sidereal
time bin $k$.  The function $R(\alpha - \tau, \delta)$ is the
differential event rate for an isotropic cosmic ray flux in the
detector coordinate system (parameterized by hour angle and
declination).  The form of this function is determined by the
overburden and the zenith angle dependence of the cosmic ray flux.  The
function $\epsilon(\alpha, \delta)$ represents the true cosmic ray
anisotropy.

As stated in the previous section, the term $T_j$ in
Eqn.~\ref{eqn:n_alpha_delta} varies by $^{+1.4\%}_{-1.3\%}$.  This
variation of purely instrumental origin was removed by multiplying
each sidereal time bin by the following weight:

\begin{equation}
\label{eqn:exposure_weight}
w_k = \frac{\left<T\right>}{T_k}
\end{equation}

\noindent Once this correction is made, the raw anisotropy data can be
expressed as follows:

\begin{eqnarray}
\label{eqn:small_n_alpha_delta}
n(\alpha_i, \delta_j) & = & \sum_{k}^{N_{sid}} \left[ 1 +
  \epsilon(\alpha_i, \delta_j) \right] \cdot R(\alpha_i - \tau_k,
  \delta_j) \\
\nonumber
 & = & \left[ 1 + \epsilon(\alpha_i, \delta_j) \right] \cdot
  \rho(\delta_j)
\end{eqnarray}

\noindent In the second line, $\rho(\delta_j) = \sum_{k}^{N_{sid}}
R(\alpha_i - \tau_k, \delta_j)$ -- i.e. the summation of hour angle
erases the right ascension dependence because of the cyclical nature
of the function $R$.  In other words, the sum is independent of the
starting point, specified by $\alpha_i$.

Ideally, one would like to extract the anisotropy function $\epsilon(\alpha, \delta)$ from the data.  In practice, this cannot be done because the declination dependence for an isotropic flux, $\rho(\delta)$, can neither be measured from the data nor calculated to an accuracy required to extract an anisotropy of order several parts per 10,000.  However, $\rho(\delta)$ can be factored out by calculating the following ratio:

\begin{eqnarray}
\label{eqn:a_alpha_delta}
A(\alpha_i, \delta_j) & = & \frac{n(\alpha_i, \delta_j) - \left<
  n(\delta_j) \right>}{\left< n(\delta_j) \right>} \\
\label{eqn:a_alpha_delta_2}
 & \approx & \epsilon(\alpha_i, \delta_j) - \left< \epsilon(\delta_j)
  \right>
\end{eqnarray}

\noindent The second line is an approximation that ignores second and
higher order terms in $\epsilon$.  The averages indicated by the
brackets is over right ascension bins.  Explicitly,

\begin{eqnarray}
\label{eqn:navg_vs_delta}
\left< n(\delta_j) \right> & = & \frac{1}{N_\alpha} \cdot
\sum_{i}^{N_\alpha} n(\alpha_i, \delta_j) \\
\nonumber
 & = & \left[ 1 + \left< \epsilon(\delta_j) \right> \right] \cdot
\rho(\delta_j)
\end{eqnarray}

\begin{equation}
\label{eqn:epsavg_vs_delta}
\left< \epsilon(\delta_j) \right> = \frac{1}{N_{\alpha}}
\sum_{i}^{N_{\alpha}} \epsilon(\alpha_i, \delta_j)
\end{equation}

The second term in Eqn.~\ref{eqn:a_alpha_delta_2} distorts the true anisotropy function $\epsilon(\alpha, \delta)$ in an unknown, but restricted way.  As an illustration of the nature of the distortion, let us imagine that, for a fixed declination $\delta$, $\epsilon(\alpha, \delta)$ is well-described by a sinusoidal function; this function can be written as a sum of a constant offset $\left<\epsilon(\delta)\right>$ and a sinusoidal term whose average is zero.  The second term in Eqn.~\ref{eqn:a_alpha_delta_2} removes the constant offset.  In more precise mathematical terms, this distortion can be described as follows.  The spherical harmonic decomposition of $\epsilon(\alpha, \delta)$ is:

\begin{equation}
\epsilon(\alpha, \delta) = \sum_{\ell,m} a_{\ell,m} \cdot
Y_{\ell,m}(\alpha, \lambda)
\end{equation}

\noindent The angle $\lambda = \pi/2 - \delta$ is the complement of
the declination, and it is measured relative to the $z$ axis (Earth's rotation axis) in the
usual notation for spherical harmonics.  The modified anisotropy function
$A(\alpha, \delta)$ is related to this as follows:

\begin{eqnarray}
\nonumber
A(\alpha, \delta) & = & \sum_{\ell,m} a_{\ell,m} \cdot \left[
  Y_{\ell,m}(\alpha, \lambda) - \frac{1}{2\pi} \int d\alpha^{\prime}
 \; Y_{\ell,m}(\alpha^{\prime}, \lambda) \right] \\
 & = & \sum_{\ell,m} b_{\ell,m} \cdot Y_{\ell,m}(\alpha, \lambda)
\end{eqnarray}

\noindent The new coefficients $b_{\ell,m}$ have the following values:

\begin{equation}
b_{\ell,m} = \left\{ \begin{array}{ll}
0 & m = 0 \\
a_{\ell,m} & m \neq 0
\end{array}
\right.
\end{equation}

\noindent It is seen that the axisymmetric terms (i.e. $m = 0$ terms) are zeroed out.

As a concrete example, consider the effect of the distortion on the first harmonic
of an axisymmetric anisotropy (i.e. a dipole anisotropy) of magnitude
$D$ along an arbitrary direction $(\alpha_0, \delta_0)$ in equatorial
coordinates.  The anisotropy function has the following form:

\begin{equation}
A(\alpha,\delta) = D \left[ \cos\delta \cos\delta_0 \cos(\alpha -
  \alpha_0) + \sin\delta \sin\delta_0 \right]
\end{equation}

\noindent After subtracting the constant offset, this becomes:

\begin{eqnarray}
\tilde{A}(\alpha, \delta) & = & D \cos\delta_0 \cos\delta
\cos(\alpha - \alpha_0) \\
\label{eqn:dipole_proj}
 & = & \tilde{D} \cos\delta \cos(\alpha - \alpha_0)
\end{eqnarray}

\noindent The second line is the form of the anisotropy for the
projection of the original dipole in the equatorial plane; the dipole
strength $\tilde{D} = D \cos\delta_0$ is the length of this
projection.

Functions with higher harmonics behave in more complicated ways.  For instance, consider an anisotropy function that can be described with two cones, one with excess flux, the other with a deficit flux.  If these two cones are not 180$^\circ$ opposite one another, significant contributions from higher harmonics must, by necessity, be present.  If the declination of the two cones is similar, then the distortion is small (the excess cone cancels out the deficit cone, making the constant offset small).  If they are different, the distortion pushes the declination of the two cones towards each other, while the right ascension is not affected at all.

\subsection{Analysis Method: Implementation}

In the analysis presented here, the cosmic ray anisotropy in the
celestial sphere was plotted in one and two dimensions.  The two
dimensional map corresponds to $A(\alpha, \delta)$ shown in
Eqn.~\ref{eqn:a_alpha_delta}.  The one dimensional map can be plotted
either as a function of right ascension or sidereal time.  In the
former, the plot corresponds to the following:

\begin{equation}
a(\alpha_i) =  \frac{m(\alpha_i) - \left< m \right>}{\left< m
  \right>},
\end{equation}

\noindent where $m(\alpha)$ is defined as:

\begin{equation}
m(\alpha_i) = \sum_{j}^{N_{\delta}} n(\alpha_i, \delta_j),
\end{equation}

\noindent $\left< m \right>$ corresponds to $m(\alpha)$ averaged over
right ascension, and the function $n(\alpha, \delta)$ is defined in
Eqn.~\ref{eqn:small_n_alpha_delta}.  The one dimensional map as a
function of sidereal time is defined as follows:

\begin{equation}
\tilde{a}(\tau_k) = \frac{\tilde{m}(\tau_k) - \left< \tilde{m} \right>}{\left<
  \tilde{m} \right>}
\end{equation}

\noindent The function $\tilde{m}(\tau_k)$ is defined as:

\begin{equation}
\tilde{m}(\tau_k) = \sum_{i}^{N_\alpha} \sum_{j}^{N_\delta} \left[ 1 +
  \epsilon(\alpha_i, \delta_j) \right] \cdot R(\alpha_i - \tau_k,
  \delta_j),
\end{equation}

\noindent and $\left< \tilde{m} \right>$ is the average of $m(\tau)$
over sidereal time.

In practice, the one dimensional anisotropy plot $a(\alpha)$ is made
by first making a histogram of the muon track right ascension, where
each entry is weighted by $w_k$ in Eqn.~\ref{eqn:exposure_weight} in
order to equalize the exposure to all directions in the celestial
sphere (the value of the weight is $1 \pm \epsilon$, with the
correction $\epsilon$ about 1\%).  The relative variation of this
histogram about its mean corresponds to $a(\alpha)$.  The two
dimensional anisotropy $A(\alpha, \delta)$ is made exactly like
$a(\alpha)$, but in $10^\circ$ strips of declination.  Finally, the
one dimensional plot $\tilde{a}(\tau)$ is made by making a histogram
of the number of muon events in sidereal time bins, dividing this
bin-by-bin with a histogram of the detector live time in sidereal time
bins, and taking the variation relative to the mean.  The function
$\tilde{a}(\tau)$ can also be thought of as $a(\alpha)$ with $\alpha$
replaced with $\tau$.  The shape of the resulting function is similar
to that of $a(\alpha)$ because $R(\alpha - \tau, \delta)$ is generally
largest when $\alpha \approx \tau$ (Fig.~\ref{fig:overburden_norm}).
In other words, at any given moment, the right ascension of a muon
track measured with SK is approximately equal to the sidereal time, or
to the right ascension of the zenith.  For this reason, we shall refer
to $\tilde{a}(\tau)$ as the `zenith-type' anisotropy, while
$a(\alpha)$ shall be referred to as the `track-type' anisotropy
because it is made using information from muon tracks.  The
`zenith-type' anisotropy is equivalent to smearing the `track-type'
anisotropy.  Clearly, the function $a(\alpha)$ is a better probe of
cosmic ray anisotropy than $\tilde{a}(\tau)$, but we have,
nevertheless, produced $\tilde{a}(\tau)$ because most other
underground muon measurements are presented in this way.

Spurious sidereal variation of atmospheric origin described in
Sec.~\ref{sec:detector_and_data} was subtracted from all plots and
maps unless otherwise noted.  The spurious variation has little effect
on the best fit value of the parameters describing the anisotropy, but
it significantly increases the uncertainty.

\subsection{Right Ascension Distribution}

\begin{figure}
\includegraphics[width=25em]{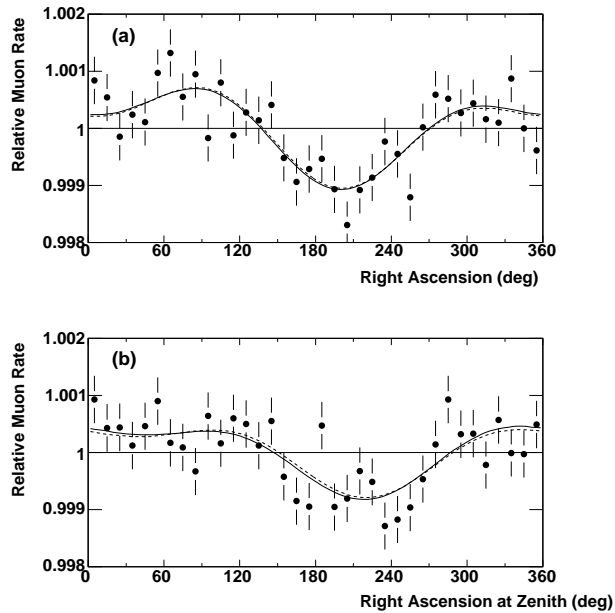}
\caption{\label{fig:1d_anisotropy_v6_deg} (a) Track-type right
  ascension projection plot.  (b) Zenith-type plot.  The error bars
  represent statistical errors.  The solid curve in each frame is the
  best fit of the first two harmonic functions.  The dashed curve
  (almost overlapping the solid curve) is the first two harmonics
  after subtracting the atmospheric contribution.}
\end{figure}

A track-type plot of the right ascension of cosmic rays before
subtracting the spurious sidereal anisotropy from atmospheric effects
is shown as data points in Fig.~\ref{fig:1d_anisotropy_v6_deg}~(a).
The solid curve is the best fit of the first two harmonics to the
data, while the dashed curve (almost overlapping the solid one) is the
sidereal variation after correcting for the atmospheric effect.  The
curves are parameterized as follows:

\begin{equation}
\label{eqn:harm2}
F(x) = A_1 \cdot \cos \left[ \frac{\pi}{180} \cdot (x - \phi_1)
  \right] + A_2 \cdot \cos \left[ \frac{2\pi}{180} \cdot (x - \phi_2)
  \right]
\end{equation}

\noindent The best fit parameters are summarized in
Table~\ref{tab:1d_aniso_best_fit_params}.  The parameter errors are
statistical, except measurements with atmospheric correction ({\sc
track/corr.} and {\sc zenith/corr.}), where the first error is
statistical and the second error is the systematic error introduced by
subtraction of the atmospheric effect (see
Appendix~\ref{app:atmospheric_effects}).


The zenith-type plot of cosmic ray right ascension is shown in
Fig.~\ref{fig:1d_anisotropy_v6_deg}~(b).  The fit parameters of (b)
are similar to those of (a), but the amplitudes are smaller, which is
consistent with the fact that (b) is obtained by smearing (a).

Anisotropy measurements based on zenith-type plots from
Kamiokande~\cite{Kamiokande} and MACRO~\cite{Macro} (both deep
underground experiments like SK-I) are also summarized in
Table~\ref{tab:1d_aniso_best_fit_params}.  Only the first term in
Eqn.~\ref{eqn:harm2} was fit to their data.  The amplitude and phase
show good agreement with those of SK-I.

\begin{figure*}
\includegraphics[width=55em]{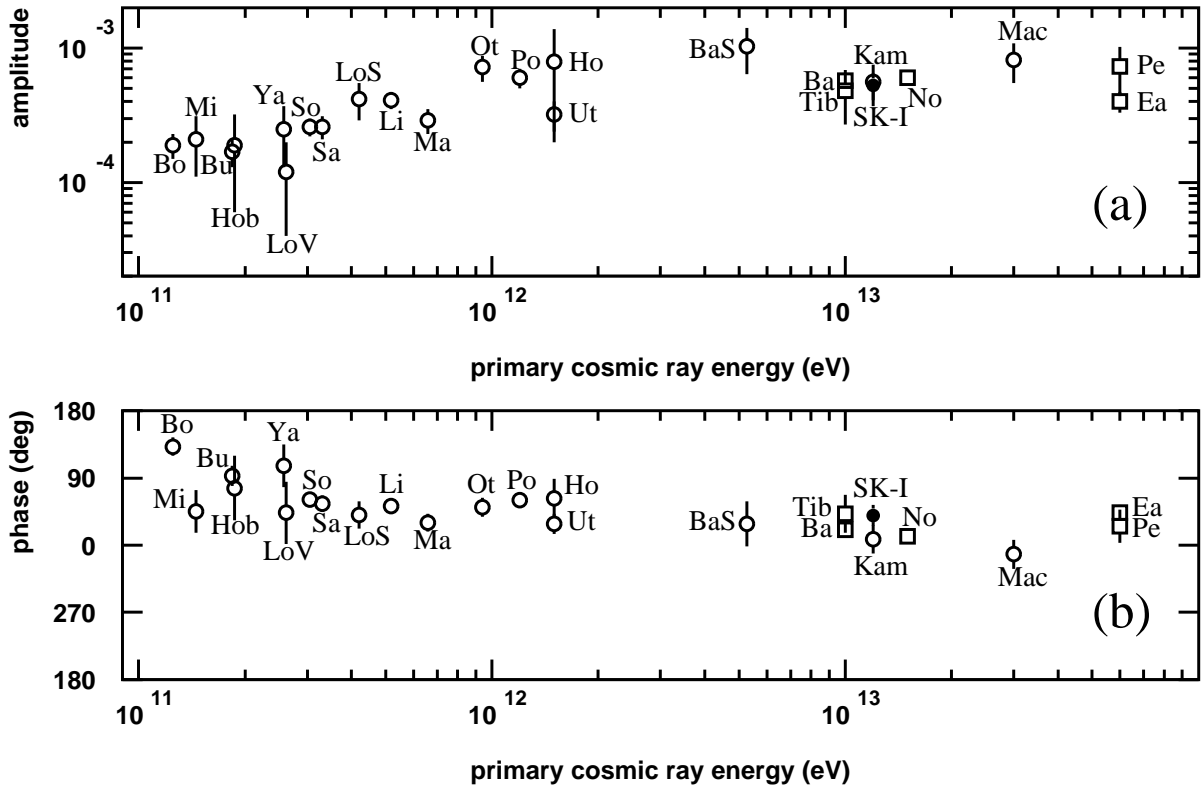}
\vspace{-45ex}
\caption{\label{fig:amp_phase_vs_energy} Amplitude and phase of the
  first harmonic fit to zenith-type plots from various cosmic ray
  experiments.  The energy in the horizontal axis is either the median
  or the log-mean energy.  Circles: muon detectors.  Squares:
  extensive air shower array.  Filled circle: SK-I.  Data references:
  Bo = Bolivia (vertical)~\cite{SWINSON_AND_NAGASHIMA}, Mi = Misato
  (vertical)~\cite{NAGASHIMA_1985}, Bu =
  Budapest~\cite{NAGASHIMA_1985}, Hob = Hobart
  (vertical)~\cite{NAGASHIMA_1985}, Ya =
  Yakutsk~\cite{NAGASHIMA_1985}, LoV = London
  (vertical)~\cite{NAGASHIMA_1985}, So = Socomo
  (vertical)~\cite{SWINSON_AND_NAGASHIMA}, Sa = Sakashita
  (vertical)~\cite{Sakashita}, LoS = London (south)~\cite{London}, Li
  = Liapootah (vertical)~\cite{Liapootah}, Ma = Matsushiro
  (vertical)~\cite{Matsushiro}, Ot = Ottawa (south):~\cite{Ottawa}, Po
  = Poatina (vertical)~\cite{Poatina}, Ho = Hong Kong~\cite{HK}, Ut =
  Utah~\cite{Mayflower}, BaS = Baksan (south)~\cite{Baksan}, SK-I
  (this report), Kam = Kamiokande~\cite{Kamiokande}, Mac =
  MACRO~\cite{Macro}, Tib = Tibet (vertical)~\cite{Tibet}, Ba = Baksan
  air shower~\cite{Baksan_EAS}, No =
  Mt. Norikura~\cite{NAGASHIMA_1989}, Ea = EAS-TOP~\cite{EASTOP1}, Pe
  = Peak Musala~\cite{Gombosi}.}
\end{figure*}

Figure~\ref{fig:amp_phase_vs_energy} shows the amplitude and phase of
the best-fit first harmonic function fit to zenith-type plots from
numerous experiments.  The SK-I result is consistent with the trend.

\begin{table*}
\begin{tabular}{|c|c|c|c|c|c|c|c|c|c|c|} \hline\hline
1 & 2 & 3 & 4 & 5 & 6 & 7 & 8 & 9 & 10 & 11 \\ \hline
{\sc exp} & {\sc plot type} & {\sc area} & {\sc depth} & {\sc lt} &
  {\sc events} & $A_1$ & $\phi_1$ &
  $A_2$ & $\phi_2$ & $\chi^2$/dof \\
 & & (m$^2$) & (m.w.e.) & (days) &  $(\times 10^6)$ & $(\times
  10^{-4})$ &  {\sc (deg)} & $(\times 10^{-4})$ &  {\sc (deg)} & \\
  \hline\hline
{\sc sk-i} & {\scriptsize \sc track/corr.} & $\approx 1000$ & $\sim
  2700$ & 1662 & 210 & $6.6 \pm 1.0 \pm 0.6$ & $33 \pm 8 \pm 5$ & $4.1
  \pm 1.0$ & $106 \pm 7$ & --- \\ \cline{2-2}\cline{7-8}\cline{11-11}
 & {\scriptsize\sc track/unc.} & & & & & $6.8 \pm 1.0$ & $30 \pm 8$ &
  & & 35.1/32 \\ \cline{2-2}\cline{7-11}
 & {\scriptsize\sc zenith/corr.} & & & & & $5.3 \pm 1.0 \pm 0.7$ & $40
  \pm 10 \pm 10$ & $2.6 \pm 1.0$ & $130 \pm 11$ & --- \\
  \cline{2-2}\cline{7-8}\cline{11-11}
 & {\scriptsize\sc zenith/unc.} & & & & & $5.7 \pm 1.0$ & $35 \pm 10$ & & &
 38.5/32 \\ \cline{1-3}\cline{5-11}
{\sc kam} & {\scriptsize\sc zenith} & $\approx 150$ & & 2072 & 59 &
  $5.6 \pm 1.9$ & $8 \pm 19$ & --- & --- & $0.3/6$ \\ \hline
{\sc mac} & {\scriptsize\sc zenith} & $\approx 1000$ & $\sim 3800$ &
  2145 & 44 & $8.2 \pm 2.7$ & $-12 \pm 20$ & --- & --- & $4.6/5$ \\
  \hline\hline
\end{tabular}
\caption{\label{tab:1d_aniso_best_fit_params} Summary of one
  dimensional anisotropy measurements from deep underground muon
  telescopes.  Col. 1: {\sc sk-i}, {\sc kam}, and {\sc mac} refer to
  the SK-I, Kamiokande~\cite{Kamiokande}, and MACRO~\cite{Macro}
  experiments.  Col. 2: type of plot.  {\sc track} = track-type plot,
  {\sc zenith} = zenith-type.  {\sc corr.} = plot corrected for
  spurious sidereal anisotropy of atmospheric origin, {\sc unc.} =
  plot uncorrected for this.  Col. 3: nominal value of detector
  projected area, in m$^2$.  Col. 4: nominal value of the overburden,
  in m.w.e..  Col. 5: total detector live time, in days.  Col. 6:
  total number (millions) of events.  Cols. 7-10: Best fit first and
  second harmonic amplitude and phase.  Errors are statistical except
  entries with two errors, where the first error is statistical and
  the second is the systematic error introduced in subtracting the
  atmospheric effect.  Col. 11: $\chi^2$ per degree of freedom of fit
  of Eqn.~\ref{eqn:harm2} to the data ($\chi2$ does not apply to data
  corrected for the atmospheric effect).  The second harmonic
  amplitude and phase are the same for the corrected and uncorrected
  result because the spurious atmospheric anisotropy is assumed to
  vary as a first harmonic function.  Kamiokande and MACRO report only
  a first harmonic fit to their data.}
\end{table*}

\subsection{Sky Map of the Anisotropy}

\begin{figure*}
\includegraphics[width=\textwidth]{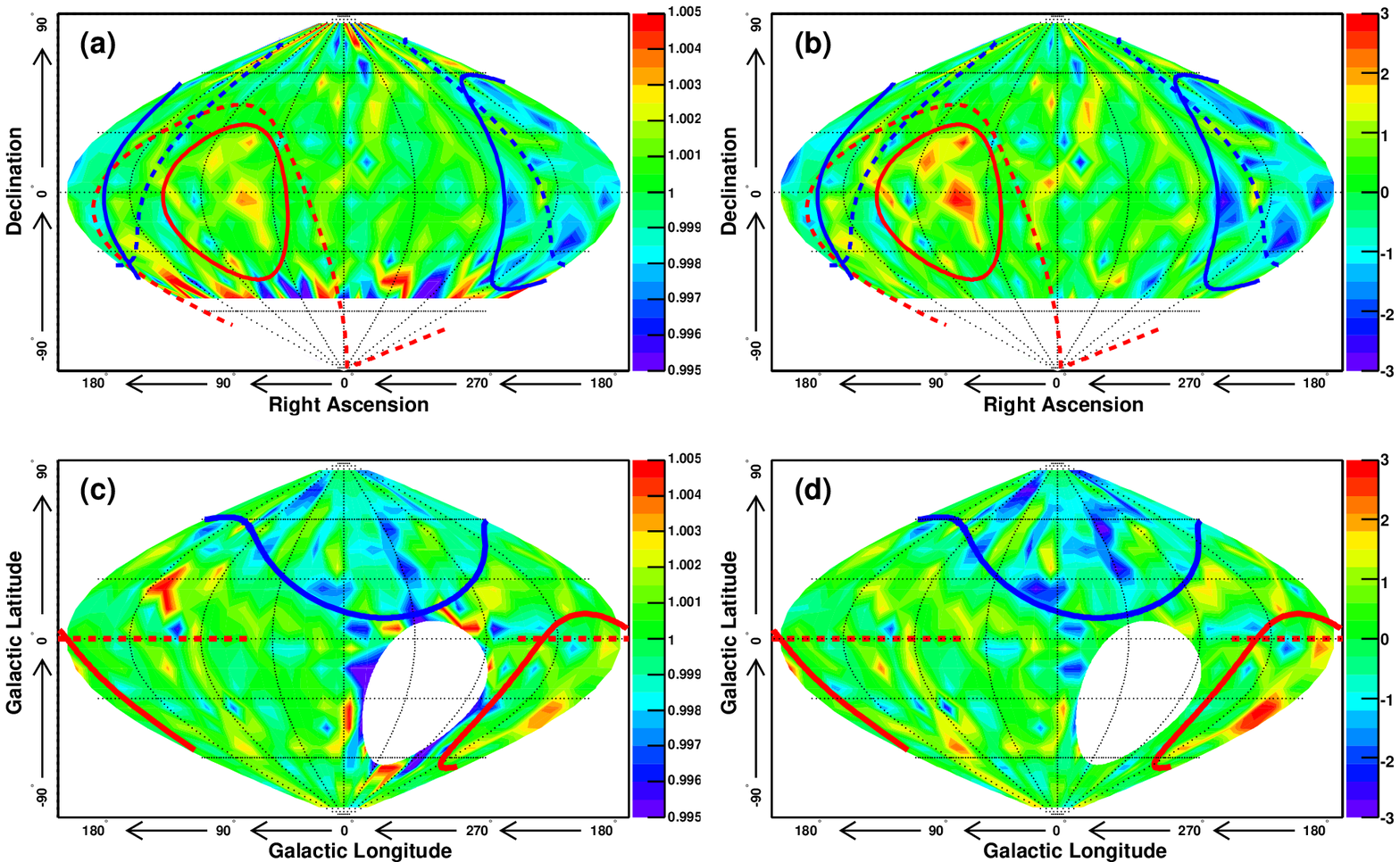}
\caption{\label{fig:2d_anisotropy_eq_sinproj} Sky map of the
  anisotropy in equatorial coordinates.  The sky is divided into
  $10^\circ \times 10^\circ$ cells (Gouraud smoothing applied only for visual purposes).
  Declinations less than $-53.58^\circ$ (white region) always lie
  below the horizon and are thus invisible to the detector.  In (a),
  each cell shows the fractional variation from the isotropic flux,
  while in (b) it shows the standard deviation of this variation.  The
  solid red and blue curves show the excess and deficit cones obtained
  using a clustering algorithm applied to the data.  The dashed curves
  in (a,b) show excess and deficit cones from the NFJ
  model~\cite{NFJ}, which is described in
  Section~\ref{sec:discussion}.  (c) and (d) show the maps in (a) and (b) transformed to the galactic coordinates.  The solid red and blue curves are the same cones as described above.  The dashed red horizontal line indicates the direction of the Orion arm.  The white patch indicates the below-horizon region.}
\end{figure*}

An anisotropy map in the celestial sphere is obtained by making
track-type plots in 10$^\circ$ strips of declination, giving $10^\circ
\times 10^\circ$ pixels.  The result is shown in
Fig.~\ref{fig:2d_anisotropy_eq_sinproj} (smoothing applied for visual
purposes only).  Each $10^\circ \times 10^\circ$ cell in map (a) shows
the fractional variation from the isotropic case, while that in (b)
shows the standard deviation of this variation.  The fractional
variations from isotropy are large and erratic in (a) because of
oversampling, but the positive and negative variations are clearly
clustered.  In order to optimize the binning, as well as to identify
and characterize the excess and deficit regions, we used a clustering
algorithm described in Appendix~\ref{app:clustering_algorithm}.  The
algorithm indicates a unique region of excess in the direction of the
constellation Taurus ($\alpha_{T}, \delta_{T}$) = ($75^\circ \pm
7^\circ$, $-5^\circ \pm 9^\circ$), while a unique region of deficit
was found in the constellation of Virgo ($\alpha_{V}, \delta_{V}$) =
($205^\circ \pm 7^\circ$, $5^\circ \pm 10^\circ$).  The half opening
angle of the ``Taurus'' region is $39^\circ \pm 7^\circ$ with a
relative rate $(0.104 \pm 0.020)$\% above average, while the size of
the ``Virgo'' region is $54^\circ \pm 7^\circ$ with a relative rate of
$(0.094 \pm 0.014)$\% below average.  All errors are statistical (see
Appendix~\ref{app:clustering_algorithm} for the method used to obtain
the statistical error of reconstructed cone parameters).  These
results are summarized in Table~\ref{tab:cone_params}.

\begin{table*}
\begin{tabular}{|c|c|c|c|c|c|} \hline\hline
1 & 2 & 3 & 4 & 5 & 6 \\ \hline
{\sc region type} & {\sc name} & {\sc cone source} & $(\alpha,
  \delta)$ & {\sc size} & {\sc deviation} \\ \hline\hline
{\sc excess} & {\sc taurus} & {\sc observed, corr.} & $(75^\circ,
  -5^\circ)$ & $39^\circ$ & ($0.104 \pm 0.020)$\% \\ \cline{3-6}
 & & {\sc observed, unc.} & $(65^\circ, 5^\circ)$ & $27^\circ$ &
  ($0.140 \pm 0.026$)\% \\ \cline{2-6}
 & {\sc tail-in} & {\sc nfj model} & $(90^\circ,
  -24^\circ)$ & $68^\circ$ & --- \\ \hline\hline

{\sc deficit} & {\sc virgo} & {\sc observed, corr.} & $(205^\circ,
 5^\circ)$ & $54^\circ$ & ($-0.094 \pm 0.014$)\% \\
 \cline{3-6}
 & & {\sc observed, unc.} & $(205^\circ, 5^\circ)$ & $54^\circ$ &
 ($-0.099 \pm 0.014$)\% \\ \cline{2-6}
 & {\sc galactic} & {\sc nfj model} & $(180^\circ,
 20^\circ)$ & $57^\circ$ & --- \\ \hline\hline

\end{tabular}
\caption{\label{tab:cone_params} Cone parameters of excess and deficit
  regions.  The NFJ model is described in
  Section~\ref{sec:discussion}.  Column 3 describes whether the cone is from observation or model, and whether or not the atmospheric correction has been applied.  Columns 4 and 5 show the center and
  half opening angle of the cones.  Column 6 shows the deviation from
  the isotropic event rate in the contained regions (the error is the
  statistical error based on the number of events contained in the
  cone).  Rows labeled ``{\sc corr.}'' and ``{\sc unc.}'' refer,
  respectively, to cones obtained from the anisotropy map corrected
  and uncorrected for atmospheric effects.}
\end{table*}

The observed anisotropy is unlikely to be due to a random fluctuation
of an isotropic cosmic ray flux.  The calculation of the statistical
significance and the rejection of the null hypothesis are performed as
follows.  The number of events in the Taurus excess cone is $(0.104
\pm 0.020)$\% above the expectation from the isotropic distribution, which
corresponds to a gaussian probability of $2.0\times 10^{-7}$.  However, since
the entire sky above the horizon was searched with a variable half-opening angle, the actual probability for this sort of deviation to occur is larger by some trials factor.  In order to determine this, $1 \times 10^{7}$ isotropic sky maps were generated with statistical fluctuations generated with a random number generator.  To cover the angular size of the Taurus excess, we counted the number of maps with reconstructed cone with in-cone standard deviation $> 5.2$~sigma and half-opening angle between $30^\circ$ and $60^\circ$.  The number of such maps was 378 out of $10^{7}$ generated maps, giving a post-trials probability of $3.78 \times 10^{-5}$.  Similarly, the number of events in
the Virgo deficit cone is $(0.094\pm 0.014)$\% below the isotropic
expectation, which is a 6.7 standard deviation effect corresponding to a gaussian probability of $2.1\times 10^{-11}$.  Among the $1 \times 10^{7}$ generated maps, none had a deviation as large as observed with half-angle between $30^\circ$ and $60^\circ$.  We, therefore, set a 90\% confidence level upper limit of the post-trials probability at $2.3 \times 10^{-7}$.

Finally, we note that comparison of the averages between different
declination bands are not meaningful; the above analysis is,
therefore, insensitive to the excess/deficit from the direction of the
celestial poles.  In other words, the 2-dimensional anisotropy can be
thought of as a series of 1-dimensional curves in consecutive strips
of declination.  Before going through the analysis filter, each curve
is described by a constant offset corresponding to the average flux,
and a sum of harmonics whose average is zero.  The filter removes the
constant offset, keeping all other terms intact.  Thus, the analysis
presented here is insensitive to any anisotropy along Earth's rotation
axis.

\section{The Robustness of the Observed Anisotropy}

The result of the analysis is insensitive to the exact choice of the
track length and zenith angle cuts.  As an illustration of this
insensitivity, track-type plots were made for 60 combinations of track
length and zenith angle cuts; the former was varied between 0 and
20~m, and the latter between $30^\circ$ above horizontal to $90^\circ$
below horizontal (i.e. no zenith angle cut).  The harmonic function
Eqn.~\ref{eqn:harm2} was fit to each plot, and the RMS spread for each
of the four parameters was found to be within 50\% of the statistical
error.

\begin{figure}
\includegraphics[width=27em]{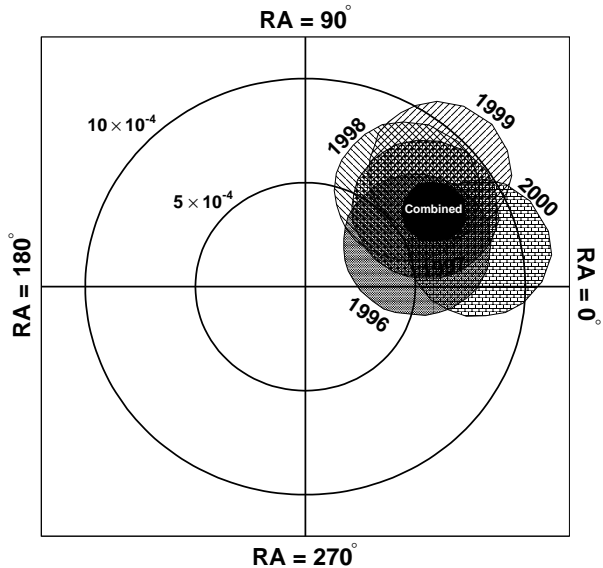}
\caption{\label{fig:amp_vs_phase_68pct_yearly_v4} The 2-parameter 68\%
confidence level regions ($\Delta \chi^2 = 2.3$) of the amplitude and
phase of the first harmonic function fit to yearly track-type
anisotropy plots.  The radial distance from the origin is the first
harmonic amplitude, while the counterclockwise angle from RA =
$0^\circ$ is the right ascension at maximum.  The regions are labeled
by the corresponding year.  The label ``Combined'' indicates the
contour from the 5-year combined data set.}
\end{figure}

As a test of signal robustness, the data were divided into five
exact-year periods spanning June~$1^{st}$ to May~$31^{st}$ of every
year from 1996 to 2000, and a measurement of anisotropy from the
track-type plot was made on each set.  The best fit first harmonic
amplitude and phase are shown in
Fig.~\ref{fig:amp_vs_phase_68pct_yearly_v4}, together with their
2-parameter 68\% confidence level regions.  Good overlap is seen.  The
fact that the phase is consistent from year to year is strong evidence
that the observed anisotropy is due to a real physical effect.

\section{Discussion}
\label{sec:discussion}

The near-isotropy of cosmic rays with energy per nucleus below the
``knee'' in the spectrum is usually described as follows.  At these
energies, strong evidence exists to indicate that primary cosmic ray
nuclei mostly originate as interstellar matter in the Milky Way
Galaxy.  They are accelerated by blast waves from supernova remnants,
and are effectively trapped in the Milky Way by the Galactic magnetic
field with a strength of order several micro-Gauss.  The gyro-radius
of a $10^{11}$~eV $\sim$ $10^{14}$~eV proton propagating in a uniform
micro-Gauss level magnetic field is in the range $\sim 10$~AU and
$\sim 0.1$~pc, much smaller than the thickness of the Galactic disk of
$200 \sim 300$~pc.  The motion of cosmic ray nuclei is spiral-like in
regions of the Galaxy where the magnetic field is smooth.
Interspersed in these regions are areas where the magnetic field is
irregular, in which the trajectories are complex and the direction
before and after entrance in these areas is nearly random.  Over large distance scales,
the irregular regions can be thought of as scatterers of cosmic rays.
As a result, the average motion of cosmic rays in the Galaxy is
expected to be highly random, which is consistent with the observed
near-isotropy of the flux.


The anisotropy at these energies is presumably due to a number of
mechanisms.  At the largest scales, the distribution of cosmic ray
sources, the large-scale configuration of the Galactic magnetic field,
the distribution of magnetic field irregularities, and the location of
the solar system in the Galaxy are all expected to be contributing
factors.  At smaller scales, the magnetic field configuration in the
neighborhood of the solar system and the distribution of the nearest
cosmic ray sources should contribute.  At the smallest scales, the
solar magnetic field can be ruled out at these energies, although
there are suggestions that the heliosphere may play a role at energies
around $\sim$~1~TeV (e.g. \cite{NFJ}).  Of course, the Compton-Getting
effect is expected to produce a dipole anisotropy on top of all of the
above.  The relative importance of each of these effects is not known.

As stated in section~\ref{sec:celestial}, Earth-based cosmic ray
anisotropy measurements at these energies require the application of a
filter to the data in order to remove the large uncertainty in zenith
angle dependence of the cosmic ray flux; this filter also removes the
axisymmetric component of the anisotropy along Earth's rotation axis.
The part of the anisotropy that comes through this filter is robustly
established by several experiments.  The earliest map of the large
scale anisotropy, referred to as the {\it NFJ model}, was made by
Nagashima, Fujimoto, and Jacklyn by combining data from several
different experiments in the northern and southern hemispheres~\cite{NFJ}.  The excess and deficit cones were
obtained by interpolating between one dimensional anisotropy
measurements made in several different declination strips.  Because
the data used were from very different detector types (shallow underground muon telescopes vs. surface air shower array) with
correspondingly large spread in energy sensitivity and very different
systematic uncertainties, the result was qualitative in nature.  More
recently, large detectors with correspondingly large statistics and
good pointing accuracy have come on line, and each one is able to make
a map of the large scale anisotropy.  We, as well as the Tibet air
shower experiment~\cite{TibetGamma}, have published such maps, and both agree well with
each other, as well as with that of~\cite{NFJ}.

A unique interpretation of the observed anisotropy is not possible, but it is useful to categorize the interpretations into two classes: (1) the true anisotropy is dominated by the dipole term, and (2) the higher harmonics are not negligible.  If scenario (1) were true, then the distortion introduced by the analysis method projects the dipole onto the equatorial plane.  Also, an excess cone and deficit cone should exist in $180^\circ$ opposition to each other.  The cones found in this analysis are both centered close to the equatorial plane; however, they are separated in right ascension by $130^\circ$, which is in  apparent contradiction to the dipole-dominant hypothesis.  Quantitatively, a $\chi^2$ fit of the data to an equatorial dipole gives a 13\% chance that the observed map is consistent with a pure equatorial dipole hypothesis.  Thus the data do not rule out this scenario.

Before discussing scenario (2), let us consider the implications of this scenario.  Considering the complicated nature of the origins and propagation of cosmic rays, it is reasonable to assume that several different mechanisms contribute to the overall observed anisotropy; each mechanism contributes a dipole term to the map, and the observed dipole is a sum over all the dipoles projected onto the equatorial plane.  One component of the dipole is due to the Compton-Getting effect.  Since the nature of the other mechanisms is unknown, it is not possible to extract the Compton-Getting term.  If, however, we make the extreme interpretation that the Compton-Getting effect is the dominant term, then one can extract the equatorial projection of the relative velocity between the cosmic ray rest frame and the solar system.  The $\chi^2$ fit described above gives $\tilde{D} =
(7.7^{+1.7}_{-1.5}) \times 10^{-4}$ and $\alpha_0 = 32^\circ \pm
12^\circ$ with $\chi^2$/d.o.f. = 577/538 (see Eqn.~\ref{eqn:dipole_proj}).  This corresponds to a relative velocity of $49^{+11}_{-10}$~km/s in the direction $32^\circ \pm 12^\circ$ right
ascension.  The speed is significantly smaller than the orbital speed of the solar system around the Galaxy ($\approx 200$~km/s), while it is comparable to the relative speed of neighboring stars around the sun.  Unless there is an accidental cancellation of large dipole terms, the observed speed should be about the magnitude of the actual Compton-Getting speed.  Thus one can deduce that, very likely, cosmic rays in the neighborhood of the solar system move around the galaxy with a motion similar to stars.  In other words, the cosmic ray rest frame is dragged along with stars.

\begin{figure}
\includegraphics[width=27em]{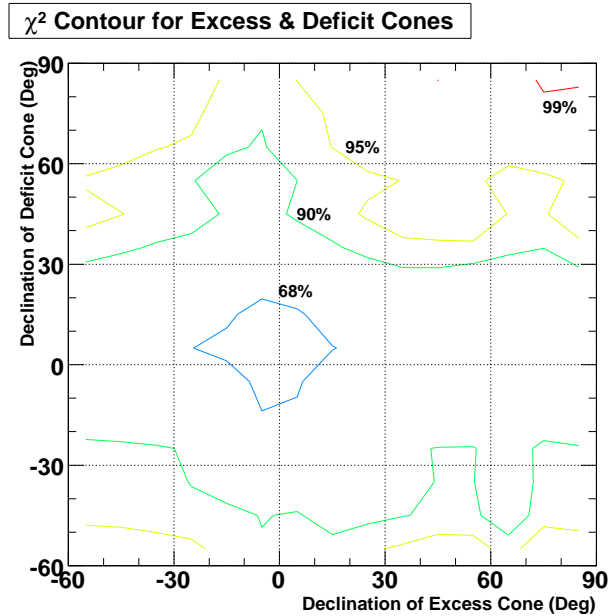}
\caption{\label{fig:marginalize_chi2} A map of the marginalized $\chi^2$ in the parameter space defined by the declination of the excess and deficit cones.  At each point in the map, $\chi^2$ was minimized with respect to the four remaining parameters.}
\end{figure}

Let us now consider scenario (2), i.e. higher harmonic terms are not negligible.  For this scenario, we focus on the particular form where two independent cones -- one with an excess flux and the other with a deficit -- produce the observed anisotropy.  The right ascension of the cone center is not affected by the distortion, whereas the declination may or may not be significantly distorted.  Specifically, if the true declination of the excess cone center is similar to that of the deficit cone, then the observed declination value should be equal to the true value (to within statistical uncertainty).  On the other hand, if there is a mismatch in the declination values, then the true values are farther apart than observed -- i.e. the filter causes the reconstructed cone declination values to get pushed towards each other.  According to Table~\ref{tab:cone_params}, the observed declination of the excess cone center is $-5^\circ$, while that of the deficit cone is $+5^\circ$; these two values are close to each other, indicating that the true values couldn't have been too far from the equator.  To put this statement on a quantitative footing, the data were compared with anisotropy maps formed with different assumptions regarding the true parameter values of the excess and deficit cones.  Each cone is defined by four parameters: the position of the cone center (two parameters), the opening angle, and the amplitude (assumed to be constant within the cone).  The right ascension of each cone was fixed to the observed value since it is not distorted by the analysis filter.  There are, thus, six free parameters that describe the two cones.  Each set of parameter values gives rise to a value of $\chi^2$ when compared against the data.  The absolute minimum $\chi^2/\mbox{d.o.f.}$ of 544/534 occurs near the set of values given in Table~\ref{tab:cone_params}.  A marginalized $\chi^2$ map is shown in Fig.~\ref{fig:marginalize_chi2}.  The figure shows $\chi^2$ as a function of declination of the excess and deficit cone center.  At each point in the map, $\chi^2$ was minimized with respect to the remaining four parameters.  The contour shows the confidence level with which any pair of declination values is allowed.  At 90\% confidence level, the declination of the excess cone is completely unconstrained.  A more stringent constraint is attainable with the deficit cone declination, but at 99\% level, it also becomes almost totally unconstrained.

Let us explore the implications of the scenario where the observed cones are close to the true ones.  A natural coordinate system for interpreting cosmic ray anisotropy in these energies is galactic.  Figs.~\ref{fig:2d_anisotropy_eq_sinproj}~(c) and (d) show the anisotropy map in galactic coordinates, (c) showing the fractional variation from isotropy, while (d) shows the standard deviation of this variation.  One feature that clearly stands out is the large deficit seen in the Galactic northern hemisphere.  This may be related to the fact that the solar system is displaced to the north of the Galactic equator by about 15~pc, a significant fraction of the half-width of the Galactic disk of about 100$\sim$200~pc.  Since the cosmic ray density is almost certainly greatest at the equator and tapering off with vertical distance away from it, the density is less to the north than to the south as viewed from the solar system.  This will tend to produce a deficit flux in the Galactic north.  Another observation in this regard is the fact that the deficit is less pronounced in the direction of the Orion arm.  The solar system is currently located at the edge of this arm and moving towards it.  It is a standard view that the cosmic ray density is elevated in the spiral arms compared to the gap regions.  Thus a density gradient is likely to produce an excess flux from the Orion arm as viewed from the solar system; this excess flux may be canceling out the deficit flux from the Galactic north.  We note that these observations essentially the same as those made in~\cite{gombosi_nature}.  A final unexplained feature is the excess cone.  We are unaware of any Galactic features that may cause this.  The  proponents of the NFJ model~\cite{NFJ}  point out that it is more-or-less aligned with the tail direction of the heliosphere; however, no plausible physical mechanism exists that could explain the size of the observed anisotropy~\cite{heliosphere_model}.

\section{Conclusion}

An anisotropy map of cosmic rays of nominal energy of 10~TeV was made
from 1662 days of observation.  The right ascension projection of this
map has a first harmonic amplitude and phase of $(6.64 \pm
0.98~\mbox{(stat.)} \pm 0.55~\mbox{(syst.)}) \times 10^{-4}$ and
$(33.2^\circ \pm 8.2^\circ~\mbox{(stat.)} \pm
5.1^\circ~\mbox{(syst.)})^\circ$, which are in good
agreement with results from other experiments.  The sky map indicates
a region with ($0.104 \pm 0.020$)\% excess flux in the constellation
of Taurus, while a region with ($0.094 \pm 0.014$)\% deficit flux is
observed in the constellation of Virgo.  The excess region is centered
at $(\alpha_T, \delta_T) = (75^\circ \pm 7^\circ, -5^\circ \pm
9^\circ)$ with a half opening angle of $39^\circ \pm 7^\circ$, while
the corresponding values for the deficit region are $(\alpha_V,
\delta_V) = (205^\circ \pm 7^\circ, 5^\circ \pm 10^\circ)$ and half
opening angle $= 54^\circ \pm 7^\circ$.  These regions largely
coincide with those of the NFJ model, and also with those observed by
the Tibet collaboration.  This agreement between experiments using very different measurement techniques spanning several decades of observation and covering primary cosmic ray energies of $\sim 1$ to $\sim 100$~TeV indicates the robustness of the observed anisotropy pattern.  The pattern, therefore, is a real feature of the cosmic ray flux in the neighborhood of the solar system at the current epoch.  The simplest model for the observation is a pure dipole pattern, which could be produced by the Compton-Getting effect, but could also be the leading harmonic term from other more complicated mechanisms.  Our observation is not described very well with a pure dipole pattern, although, at 13\% confidence level, we cannot rule it out.  The distorting effect of the filter applied to the data prevents us from making a unique physical interpretation of the observation.  If, however, it is assumed that the distortion is not too great, the deficit region coincides with a large portion of the Galactic northern hemisphere.  This may be related to the fact that the solar system is displaced by about 15~pc to the north of the Galactic plane.  Also, the deficit appears to be canceled out in the direction of the Orion Arm, which may be an indication of enhanced levels of cosmic ray density there.  The excess region does not seem to match any features that could provide a mechanism for its existence, though it has been noted~\cite{NFJ} that it points in the direction of the tail end of the heliosphere.


\begin{acknowledgments}
We gratefully acknowledge the cooperation of the Kamioka Mining and
Smelting Company.  The Super-Kamiokande experiment has been built and
operated from funding by the Japanese Ministry of Education, Culture,
Sports, Science, and Technology, the United States Department of
Energy, and the U.S. National Science Foundation.  Some of us have
been supported by funds from the Korean Research Foundation (BK21) and
Korea Science and Engineering Foundation, the Polish Committee for
Scientific Research (grant 1P03B08227), Japan Society for the
Promotion of Science, and Research Corporation's Cottrell College
Science Award.
\end{acknowledgments}


\appendix

\section{Subtracting the Anisotropy Due to Atmospheric Effects}
\label{app:atmospheric_effects}

The observed sidereal anisotropy is due to two effects:
extra-terrestrial (i.e. Compton-Getting, Galactic, and heliospheric
effects) and atmospheric (due to residual effects of seasonal and
solar diurnal variations in the atmospheric temperature).  Since we
are interested only in the extra-terrestrial anisotropy, it is
desirable to subtract the atmospheric contribution.  We discuss in
this Appendix technical details on this subtraction.  The method is
due to Farley and Storey~\cite{Farley}, which is applied to the
zenith-type plot.  In the second section this result is generalized to
the two dimensional anisotropy map.  Finally, the results of the first
two sections are used to subtract the atmospheric contribution from
the track-type one dimensional plot.

\subsection{Subtraction for the Zenith-Type Plot}
\label{subsec:rate_vs_lst}

The zenith-type plot is equivalent to the relative variation in the
muon rate as a function of local sidereal time.  In Farley and
Storey's formulation, the rate variation is parameterized generally as
follows:

\begin{eqnarray}
\label{eqn:pseudo_sidereal_1} \nonumber
R(t) & = & 1 + \overbrace{ \left[ A + 2B \, \cos 2\pi (t - \phi_2)
 \right] }^{\mbox{\sc \small seasonal modulation}} \, \overbrace{ \cos
 2\pi (Nt - \phi_1) }^{\mbox{\sc \small solar variation}} \\
 & & + \; \underbrace{ C \, \cos 2\pi \left\{ (N+1)t - \phi_3 \right\}
 }_{\mbox{\sc \small true sidereal variation}},
\end{eqnarray}

\noindent where $t$ is measured in years, $N \approx
365.24$~cycles/year is the solar diurnal frequency, and $\phi_i$, $i =
1, 2, 3$ are phase angles.  The parameters $A$, $B$, and $C$ are the
magnitude of the relative rate variation for different periodicities
(discussed below).  The solar diurnal variation is assumed to be
seasonally modulated (first line, Eqn.~\ref{eqn:pseudo_sidereal_1}).
The second line represents the true sidereal variation (i.e. of
extra-terrestrial origin).  A re-arrangement of the first line above
gives the following:

\begin{widetext}
\begin{eqnarray}
\label{eqn:rate_vs_time_alt}
R(t) & = & 1 + \overbrace{ A \, \cos 2\pi (Nt - \phi_1) }^{\mbox{\sc
 \small solar}} \\ \nonumber 
 & & + \overbrace{ B^\prime \, \cos 2\pi \left\{ (N+1)t - (\phi_1 + \phi_2)
 \right\} }^{\mbox{\sc \small spurious sidereal}} + \overbrace{ C \,
 \cos 2\pi \left\{ (N+1)t - \phi_3 \right\} }^{\mbox{\sc \small true
 sidereal}} \\ \nonumber 
 & & + \underbrace{ B \, \cos 2\pi \left\{ (N-1)t - (\phi_1 - \phi_2)
 \right\} }_{\mbox{\sc \small pseudo-sidereal}}
\end{eqnarray}
\end{widetext}

\begin{figure}
\includegraphics[width=25em]{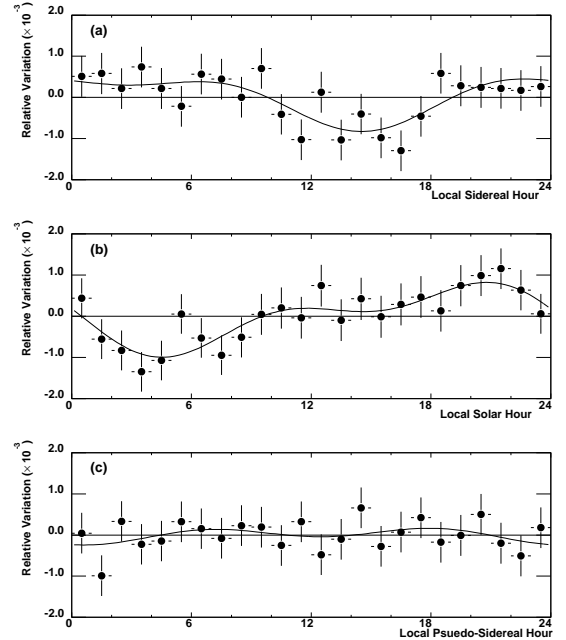}
\caption{\label{fig:1d_anisotropy_v7_hr} (a) Relative muon rate as a
  function of local sidereal time (i.e. zenith-type map), in hours
  right ascension.  (b) Relative muon rate as a function of local
  solar hour.  (c) Relative muon rate as a function of hours,
  pseudo-sidereal time.  The curve in each frame is the best fit of
  the first two harmonic functions to the data.}
\end{figure}

\noindent The first time dependent term in
Eqn.~\ref{eqn:rate_vs_time_alt} is the relative rate variation with a
periodicity of one solar day, the second and third terms are rate
variations with a periodicity of one sidereal day, and the final term
is the rate variation with a periodicity of one pseudo-sidereal day,
which is longer than the solar diurnal day by about 0.27\%
(Figs.~\ref{fig:1d_anisotropy_v7_hr}~(a)-(c)).  Written in this form,
it is seen that there are two sources of sidereal variation, the
``spurious'' one due to the atmosphere and the ``true'' one due to
extraterrestrial effects.

Each time dependent term in Eqn.~\ref{eqn:rate_vs_time_alt} can be
represented by phasors.  In Cartesian coordinates, they are given as
follows:

\begin{eqnarray}
\nonumber \vec{A} & = & \left( A \, \cos\phi_1 , \; A \, \sin\phi_1
\right) \\
\nonumber \vec{B}^{\prime} & = & \left( B^{\prime} \, \cos(\phi_1 +
\phi_2) , \; B^{\prime} \, \sin(\phi_1 + \phi_2) \right) \\
\nonumber \vec{C} & = & \left( C \, \cos\phi_3 , \; C \, \sin\phi_3
\right) \\
\nonumber \vec{B} & = & \left( B \, \cos(\phi_1 - \phi_2), \; B \, 
\sin(\phi_1 - \phi_2) \right)
\end{eqnarray}

\noindent The phasor $\vec{A}$ is non-zero due to residual effects of
the solar diurnal and seasonal variation in the atmospheric
temperature, while $\vec{D} = \vec{C} + \vec{B}^{\prime}$ is non-zero
primarily due to extra-terrestrial effects (represented by $\vec{C}$),
although, in general, a non-zero contribution is also made by
atmospheric effects (represented by $\vec{B}^{\prime}$).  No real
effect is directly responsible for a non-zero value of $\vec{B}$, but,
as described above, interplay between seasonal and solar diurnal
variation in the atmospheric temperature can indirectly give rise to a
non-zero magnitude.

\begin{figure*}
\includegraphics[width=\textwidth]{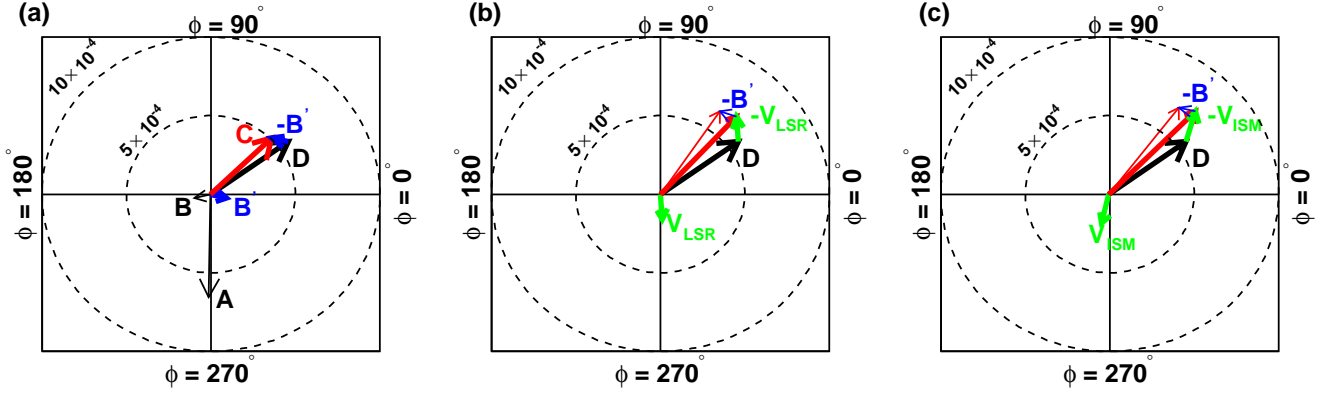}
\caption{\label{fig:phasor_demo_cg_v3} Phasor diagrams showing the
  result of subtracting the atmospheric and Compton-Getting effects.
  The length of an arrow represents the amplitude of the first
  harmonic component, while the angle measured counter clockwise from
  $\phi = 0^\circ$ is the phase at maximum.  (a) Atmospheric effect,
  (b) Compton-Getting effect assuming cosmic ray rest frame moving
  with the local standard of rest, and (c) same as (b), but moving
  with the local interstellar medium.  In each plot, the vector
  $\vec{D}$ indicates the uncorrected amplitude and phase, while the
  vector $-\vec{B^{\prime}}$, $-\vec{V}_{\mbox{\scriptsize\sc lsr}}$,
  and $-\vec{V}_{\mbox{\scriptsize\sc ism}}$ are corrections for the
  atmospheric and Compton-Getting effects.  The vectors $\vec{A}$,
  $\vec{B}$, and $\vec{C}$ are defined in the text.  (b) and (c) also
  show the result of subtracting both effects.}
\end{figure*}

In terms of phasors, one sees that the process of measuring the true
sidereal variation involves measuring the phasor $\vec{B}^{\prime}$
and subtracting it from $\vec{D}$.  The phasor $\vec{D}$ is obtained
from Fig.~\ref{fig:1d_anisotropy_v7_hr}~(a), while $\vec{B}^{\prime}$
is obtained from $\vec{B}$ and $\vec{A}$, which are, in turn, obtained
from Figs.~\ref{fig:1d_anisotropy_v7_hr}~(b) and (c).  Specifically,
$\vec{B}^{\prime}$ is obtained by reflecting $\vec{B}$ about the axis
defined by $\vec{A}$ (see Fig.~\ref{fig:phasor_demo_cg_v3}~(a)).

$\vec{B}^\prime$ and $\vec{C}$ can be obtained approximately by using
the most likely value of $\vec{A}$, $\vec{B}$, and $\vec{D}$.  These
are given as $\vec{B}^\prime$({\sc calc}) and $\vec{C}$({\sc calc}) in
Table~\ref{tab:phasor_params}.  A statistically rigorous determination
of $\vec{C}$ and associated uncertainties require the use of a
statistical subtraction technique in which the phasors $\vec{A}$,
$\vec{B}$, and $\vec{D}$ are generated randomly according to their
respective $\chi^2$ probability from the first harmonic fits in
Fig.~\ref{fig:1d_anisotropy_v7_hr}.  For each generated triplet of
phasors, there corresponds a unique value of $\vec{C}$, and an
ensemble of generated $\vec{C}$ gives a distribution of $\left|
\vec{C} \right|$ and $\phi_C$.  The mean and RMS of a Gaussian fit to
each are taken as the most likely value and uncertainty of the true
sidereal anisotropy; these are given as $\vec{C}$({\sc stat}) in
Table~\ref{tab:phasor_params}.

\begin{table}
\begin{tabular}{|c|c|c|} \hline\hline
{\sc phasor} & {\sc amp} $(\times 10^{-4})$ & {\sc phase (hr)} \\
\hline\hline
$\vec{A}$ & $6.63 \pm 0.98$ & $18.0 \pm 0.6$ \\ \hline
$\vec{B}$ & $1.01 \pm 0.98$ & $13.1 \pm 3.7$ \\ \hline
$\vec{D}$ & $5.65 \pm 0.98$ & $2.3 \pm 0.7$ \\ \hline\hline
$\vec{B}^\prime$ {\sc (calc)} & 1.01 & 22.9 \\ \hline
$\vec{C}$ {\sc (calc)} & 5.09 & 2.9 \\ \hline\hline
$\vec{B}^\prime$ {\sc (stat)} & $0.98 \pm 0.71$ & $23.2 \pm
5.2$ \\ \hline
$\vec{C}$ {\sc (stat)} & $5.31 \pm 1.19$ & $2.7 \pm 0.9$ \\
\hline\hline
\end{tabular}
\caption{\label{tab:phasor_params} Summary of phasor parameters.
  $\vec{A}$, $\vec{B}$, and $\vec{D}$ are from data (N.B. the
  amplitude and phase of $\vec{D}$ are slightly different from those
  shown in Table~\ref{tab:1d_aniso_best_fit_params} because of
  difference in binning); $\vec{B}^\prime$({\sc calc}) and
  $\vec{C}$({\sc calc}) are calculated directly from the first three,
  while $\vec{B}^\prime$({\sc stat}) and $\vec{C}$({\sc stat}) are
  obtained by the statistical subtraction technique described in the
  text.}
\end{table}

\subsection{Subtraction for the Two Dimensional Map}
\label{subsec:2d}

We describe in this section the method used to subtract the anisotropy
of atmospheric origin from the two dimensional anisotropy map.
Letting $M$ denote a two dimensional map, the true (i.e. corrected)
map $M${\scriptsize\sc true} is given by:

\begin{equation}
M\mbox{{\scriptsize\sc true}} = M\mbox{{\scriptsize\sc obs}} -
M\mbox{{\scriptsize\sc atm}},
\end{equation}

\noindent where $M$\mbox{{\scriptsize\sc obs}} is the observed map,
and $M$\mbox{{\scriptsize\sc atm}} is the map of anisotropy of
atmospheric origin.  $M$\mbox{{\scriptsize\sc atm}} is calculated by
assuming that: (1) the incident cosmic ray flux is isotropic (the
observed anisotropy introduces only a second order correction, so it
can be ignored); (2) the atmospheric effect causes the overall cosmic
ray rate to vary with amplitude and phase given by the phasor
$\vec{B}^\prime$, i.e. amplitude = $0.98 \times 10^{-4}$ and phase =
$348^\circ$ right ascension.

The map $M${\scriptsize\sc atm} is obtained by convoluting the relative
rate variation $R(\tau_s) = 1 + \left| B^{\prime} \right| \, \cos(
\omega_s \, \tau_s - \phi_{B^{\prime}})$ with the isotropic event rate
$I(\delta, h)$, where $\tau_s$ is sidereal time, $\omega_s$ is the
sidereal angular frequency, $\delta$ is the declination, $h = \alpha -
\tau_s$ is the hour angle, and $\alpha$ is the right ascension.  Note
that $I$ has units of $\mbox{day}^{-1} \; \mbox{m}^{-2} \;
\mbox{sr}^{-1}$; it is related to Fig.~\ref{fig:overburden_norm} by a
coordinate transformation.  The convolution is as follows:

\begin{equation}
\label{eqn:matm}
M\mbox{{\scriptsize\sc atm}}(\delta, \alpha) = \int \, d\tau_s \,
I(\delta, \alpha - \omega_s \tau_s) \; R(\tau_s)
\end{equation}

\begin{figure}
\includegraphics[width=25em]{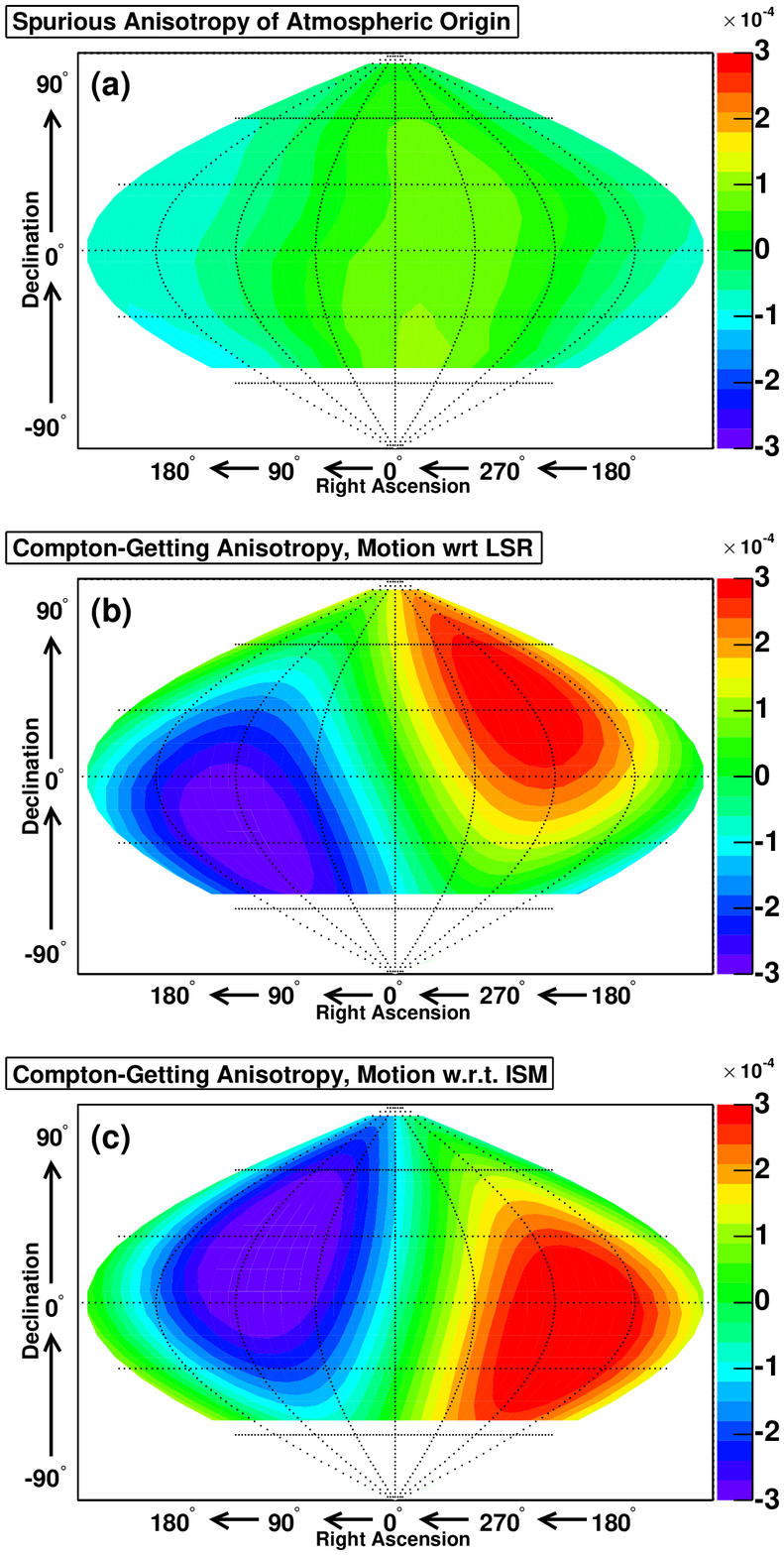}
\caption{\label{fig:2d_anisotropy_corrected} Anisotropy introduced by:
  (a) the atmospheric effect; (b) the Compton-Getting effect assuming
  that the bulk cosmic ray motion is the same as the local standard of
  rest; (c) same as (b), but the motion is assumed to be the same as
  that of the neutral interstellar matter.  The contour values
  indicate fractional deviation from isotropy.
  The white region below declination of $-53.58^\circ$ is always below
  the horizon.  Note that the filter applied to the data projects the
  anisotropy shown in (b) and (c) onto the equatorial plane.}
\end{figure}

\noindent The map $M${\scriptsize\sc atm} (from which the $\delta$
dependence is factored out) is shown in
Fig.~\ref{fig:2d_anisotropy_corrected}~(a).  The excess and deficit
cone parameters for $M${\scriptsize\sc true} are shown in
Table~\ref{tab:cone_params_cg_correct}.  The direction and cone size
of the deficit region are unchanged by this correction, whereas those
of the excess region change noticeably.

\begin{table*}
\begin{tabular}{|c|c|c|c|c|c|c|} \hline\hline
{\sc region type} & {\sc name} & {\sc cone source} & $(\alpha,
  \delta)$ & {\sc size} & {\sc deviation} & $\chi$ \\ \hline\hline 

{\sc excess} & {\sc taurus} & {\sc no correction} & $(65^\circ,
 5^\circ)$ & $27^\circ$ & 0.140\% & $5.26 \, \sigma$ \\ \cline{3-7}
 & & {\sc atm} & $(75^\circ, -5^\circ)$ & $39^\circ$ & 0.104\% & $5.31
 \, \sigma$ \\ \cline{3-7}
 & & {\sc cg, lsr} & $(85^\circ, -35^\circ)$ & $65^\circ$ & 0.0964\% &
 $6.46 \, \sigma$ \\ \cline{3-7}
 & & {\sc cg, ism} & $(55^\circ, 5^\circ)$ & $60^\circ$ & 0.0884\% &
 $6.95 \, \sigma$ \\ \cline{3-7}
 & & {\sc atm + cg, lsr} & $(85^\circ, -35^\circ)$ & $65^\circ$ &
 0.0991\% & $6.65 \, \sigma$ \\ \cline{3-7}
 & & {\sc atm + cg, ism} & $(55^\circ, 5^\circ)$ & $60^\circ$ &
 0.0877\% & $6.90 \, \sigma$ \\ \cline{2-7}
 & {\sc tail-in} & {\sc nfj model} & $(90^\circ,
 -24^\circ)$ & $68^\circ$ & --- & --- \\ \hline\hline

{\sc deficit} & {\sc virgo} & {\sc no correction} & $(205^\circ,
5^\circ)$ & $54^\circ$ & $-0.0988$\% & $-7.27 \, \sigma$ \\ \cline{3-7}
 & & {\sc atm} & $(205^\circ, 5^\circ)$ & $54^\circ$ & $-0.0940$\% &
$-6.91 \, \sigma$ \\ \cline{3-7}
 & & {\sc cg, lsr} & $(215^\circ, 5^\circ)$ & $55^\circ$ & $-0.107$\%
& $-7.95 \, \sigma$ \\ \cline{3-7}
 & & {\sc cg, ism} & $(215^\circ, 5^\circ)$ & $55^\circ$ & $-0.115$\%
& $-8.57 \, \sigma$ \\ \cline{3-7}
 & & {\sc atm + cg, lsr} & $(215^\circ, 5^\circ)$ & $55^\circ$ &
$-0.103$\% & $-7.66 \, \sigma$ \\ \cline{3-7}
 & & {\sc atm + cg, ism} & $(215^\circ, 5^\circ)$ & $55^\circ$ &
$-0.111$\% & $-8.28 \, \sigma$ \\ \cline{2-7}
 & {\sc galactic} & {\sc nfj model} & $(180^\circ,
20^\circ)$ & $57^\circ$ & --- & --- \\ \hline\hline

\end{tabular}
\caption{\label{tab:cone_params_cg_correct} Cone parameters of excess
  and deficit regions.  The column ``Cone Source'' refers to the data
  set or model from which the cones are derived.  ``No Correction''
  refers to the anisotropy without any subtraction, while the other
  rows refer to that after subtracting various effects.  ``ATM'' is
  the atmospheric effect, ``CG, LSR (ISM)'' is the equatorially
  projected Compton-Getting effect with the cosmic ray rest frame
  assumed to be the same as the local standard of rest (local
  interstellar matter).  ``ATM + CG, LSR (ISM)'' is that from which
  both effects are subtracted.  Columns 4 and 5 show the center and
  half opening angle of the cones.  Column 6 shows the deviation from
  the isotropic event rate in the contained regions, and column 7
  ($\chi$) shows the statistical significance of the deviation.}
\end{table*}

\subsection{Subtraction for the Track-Type Map}

The track-type one dimensional map corrected for atmospheric effects
can be obtained by simply projecting $M${\scriptsize\sc true}$(\delta,
\alpha)$ onto the right ascension axis.  However, this does not
provide an estimate of the uncertainty introduced by the atmospheric
subtraction.  In order to obtain this, we first generate an ensemble
of phasors $\vec{D}$, $\vec{A}$, and $\vec{B}$, as described in
Sec.~\ref{subsec:rate_vs_lst}.  For each triplet, a two dimensional
map $M${\scriptsize\sc true} is made, as described in
Sec~\ref{subsec:2d}.  This map is then projected onto right ascension
axis to obtain the one dimensional map.  The first two harmonic
functions (Eqn.~\ref{eqn:harm2}) are then fit to each map thus
obtained.  We thus obtain an ensemble of fit values $A_1$ and $\phi_1$
(the second harmonic is unchanged in the ensemble because the
atmospheric effect was assumed to have only first harmonic variation).
The result of this procedure is given in
Table~\ref{tab:1d_aniso_best_fit_params} in the row labeled ``{\sc
track/corr.}''.  The first error is statistical (from fitting harmonic
functions to the data), and the second error is from the dispersion in
the fit values obtained from the ensemble method described above.

\section{Correlation between Muon Rate and Atmospheric Temperature}
\label{app:atmospheric}

The effect of the atmosphere on the cosmic ray detection rate in
underground muon detectors is, in general, correlated with the
pressure at the detector altitude and with the atmospheric temperature
profile above the detector.  The pressure dependence becomes
unimportant compared to temperature dependence for muon threshold
energy greater than about 100~GeV~\cite{sagisaka}.

In this limit, the relative change in the muon rate with atmospheric
temperature is given by the following expression:

\begin{equation}
\label{eqn:temp_dep}
\frac{\delta I}{I} \approx \int^{x_0}_{0} \alpha(x, \overline{E}_0,
x_0) \; \delta T(x) \; dx
\end{equation}

\noindent The quantity $\alpha$ is the partial temperature
coefficient, $\delta T(x)$ is the deviation of the temperature from
the mean at atmospheric depth $x$, $\overline{E}_0$ is the threshold
muon energy, and $x_0$ is the atmospheric depth at the detector
altitude.  The mechanism for the temperature dependence of the rate is
as follows.  As the temperature rises, the atmospheric density
decreases, and the probability that a meson in a cosmic ray shower is
destroyed by interaction with air nuclei decreases.  The increased
meson mean free path implies that mesons have increased chance to
decay and produce muons.  The cosmic ray muon rate, therefore, is
positively correlated with atmospheric temperature.  The partial
temperature coefficient can be calculated numerically using inputs
such as a model atmosphere, primary cosmic ray flux, particle
production cross section, particle decay constants,
etc.\cite{sagisaka}, while $\delta T(x)$ can be obtained at discrete
atmospheric levels from meteorological measurements.  For SK-I, the
change in rate due to this effect should be $\approx \pm 1$\%, which
is more than an order of magnitude larger than the magnitude of the
sidereal anisotropy.

\begin{figure*}
\includegraphics[width=\textwidth]{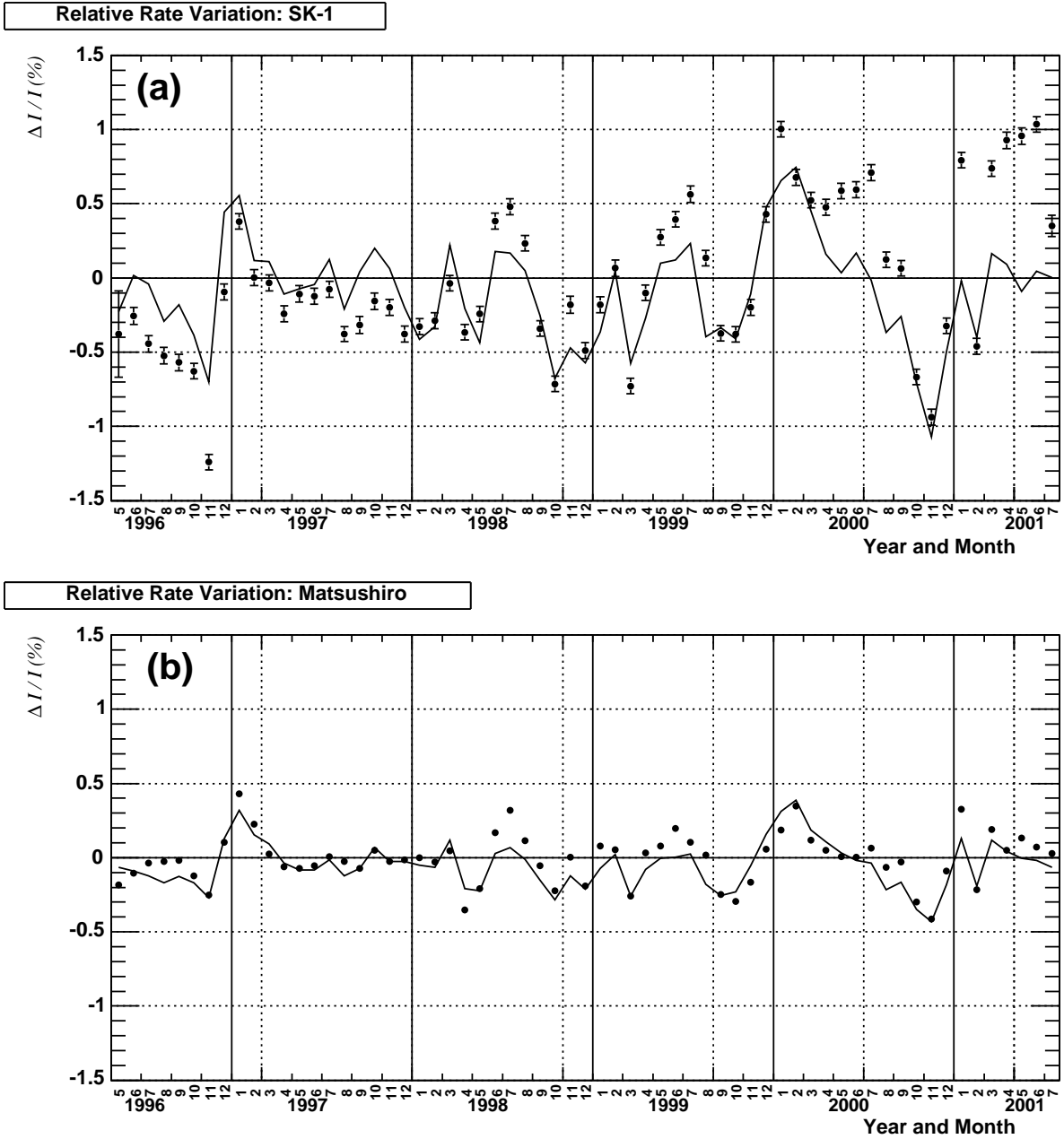}
\caption{\label{fig:muon_rate_vs_yymm_obs_and_calc} (a) Variation in
  the muon rate relative to the mean for each month of SK-I.  (b) The
  variation seen at Matsushiro during the same period.  The solid
  curve in each frame is the predicted rate variation based on
  equation~\ref{eqn:temp_dep}.}
\end{figure*}

Figure~\ref{fig:muon_rate_vs_yymm_obs_and_calc}~(a) shows the relative
variation in the muon rate for each month and year of SK-I.  The solid
curve shows the predicted variation based on
Eqn.~\ref{eqn:temp_dep}~\cite{Munakata_atm_effect_calc}.  The
temperature measurements were obtained from the Wajima Observatory
($37.38^\circ$~N, $136.90^\circ$~E, 116~km from the SK-I
detector)~\cite{Wajima}.  Radio sonde was used to measure the
temperature of 25 layers of the atmosphere between 1000~mb and 5~mb.
These measurements were made twice a day.  The agreement between the
data and prediction is good, though not perfect.  The disagreement is
due to inaccuracies in the temperature measurements at altitudes above
40~mb.  At SK-I energies, the partial temperature coefficient
increases all the way to very shallow atmospheric depths, so
inaccurate temperature measurements at altitude above 40~mb
(i.e. pressure $<40$~mb) introduce significant inaccuracies in the
predicted rate.

A better standard for checking the SK-I rate variation is by means of
a simultaneous independent measurement from a nearby underground muon
detector.  The Matsushiro detector ($36.53$~N, $138.01$~E, 79~km from
SK-I) is perfectly suited for this requirement~\cite{Matsushiro_data}.
With an overburden of 220~m.w.e., its muon energy threshold is about
100~GeV.  This lower energy threshold implies that the rate variation
at Matsushiro is similar to that of SK-I, but has smaller amplitude
(see Fig.~\ref{fig:muon_rate_vs_yymm_obs_and_calc}~(b)).

\begin{figure}
\includegraphics[width=25em]{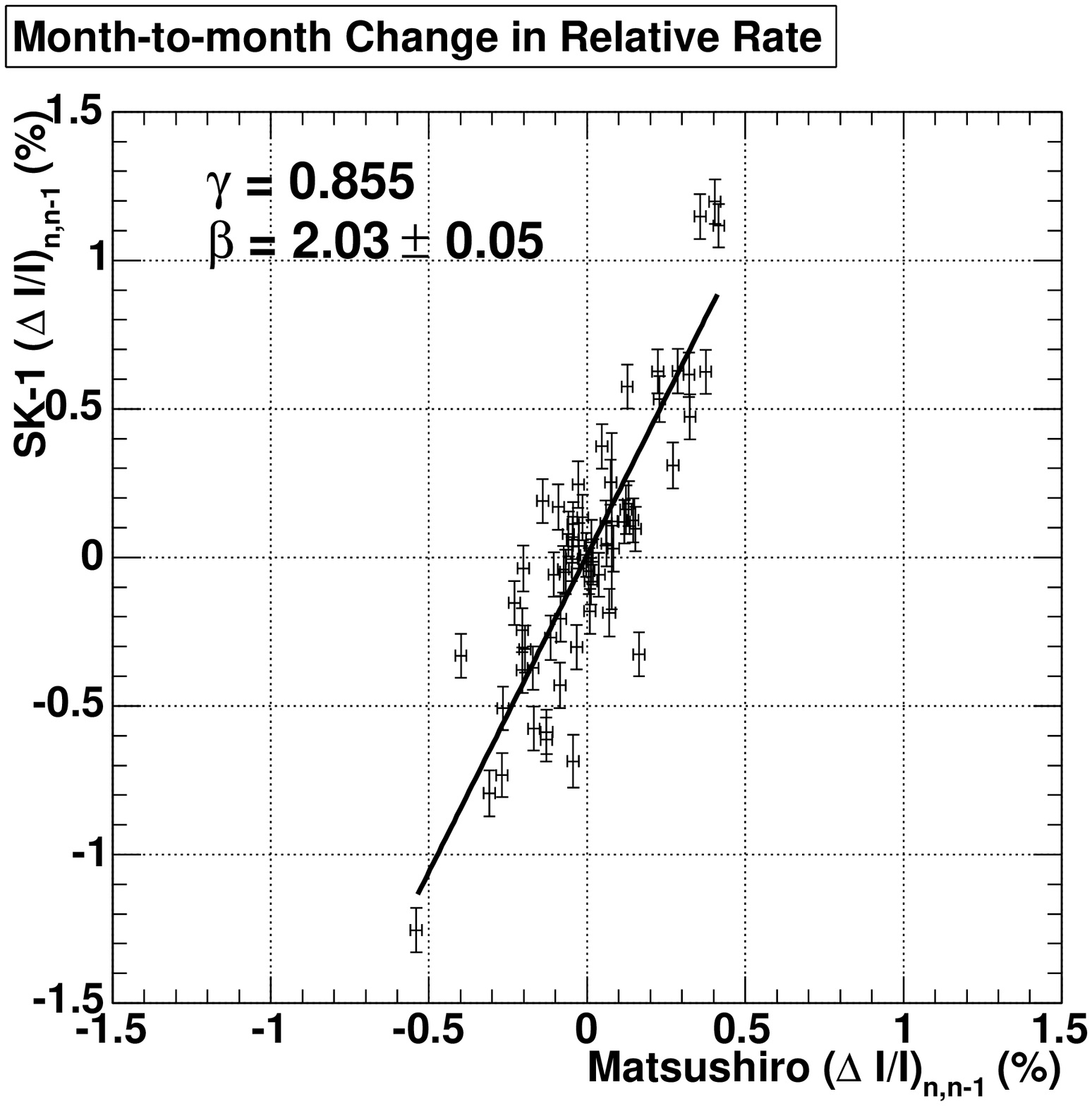}
\caption{\label{fig:change_in_rel_rate_sk_mat} The correlation
  between $(\Delta I/I)_{n,n-1}$, SK-1 {\it vs.} Matsushiro.  $\Delta
  I/I$ is the relative rate variation of a given month; the notation
  $(\Delta I/I)_{n,n-1}$ indicates the difference of this quantity
  between month $n$ and $n-1$.  $\gamma = 0.855$ is the linear
  correlation coefficient between the two data (60 degrees of
  freedom), while $\beta = 2.03 \pm 0.05$ is the best fit slope.}
\end{figure}

One measure of this difference is the month to month variation in
$\Delta I/I$ at the two sites.  According to
calculation~\cite{Munakata_atm_effect_calc}, the magnitude of this
change at SK-I should be 2.03 times larger than that at Matsushiro.  A
correlation plot of the changes at the two sites is shown in
Fig.~\ref{fig:change_in_rel_rate_sk_mat}.  The regression coefficient
$\beta = 2.03 \pm 0.05$ is in agreement with the predicted value.

When the data are binned in solar diurnal hours, the 1\% level monthly
variations almost cancel out, leaving a residual variation at the
level of several parts per ten thousand.  This variation, when
modulated seasonally, produces side band components with frequency
$365.24 \pm 1$~cycles per year.  The frequency of 366.24 cycles per
year is the inverse of one sidereal day, and the existence of this
component implies that the observed sidereal variation in the cosmic
ray rate is partly due to atmospheric temperature variations.  This
contribution to the observed sidereal anisotropy can be estimated
using the method of Farley and Storey~\cite{Farley}.  A detailed
discussion of the atmospheric contribution to the sidereal anisotropy
is given in Appendix~\ref{app:atmospheric_effects}.

\section{Subtracting the Compton-Getting Anisotropy}
\label{app:cg_correct}

The Compton-Getting effect refers to the enhancement of the cosmic ray
flux in the observer's direction of motion relative to the reference
frame in which the bulk motion of the cosmic ray plasma is at rest.
If the observer's velocity relative to the cosmic ray bulk motion is
$\vec{v}$ and the direction of observation is in the direction of the
unit vector $\hat{u}$, the relative enhancement in the intensity is
given by:

\begin{equation}
\frac{\Delta I}{I} = (2 + \gamma) \, \frac{v}{c} \, \cos \chi, 
\end{equation}

\noindent where $v = \left| \vec{v} \right|$, $c$ is the speed of
light in vacuum, and $\cos \chi = \vec{v} \cdot \hat{u} / v$ is the
cosine of the opening angle between the observer's motion and the
direction of observation.  The cosmic ray rest frame is not known,
although the observed smallness of the anisotropy implies that its
motion relative to the sun must be small ($v/c$ must be on the order
of $10^{-4}$, or $v \lesssim 30$~km/s, barring an accidental large
cancellation of the Compton-Getting anisotropy by that due to other
effects).  Two assumptions are often invoked in the literature: the
cosmic ray rest frame is at rest with respect to (1) the local
standard of rest, or (2) the local interstellar medium.  The first
motion has speed $v_{\mbox{\scriptsize\sc lsr}} \approx 20$~km/s in
the direction $(\alpha_{\mbox{\scriptsize\sc lsr}}, \,
\delta_{\mbox{\scriptsize\sc lsr}}) \approx (270^\circ, \,
29.2^\circ)$, while the second motion has speed
$v_{\mbox{\scriptsize\sc ism}} \approx 22$~km/s in the direction
$(\alpha_{\mbox{\scriptsize\sc ism}}, \, \delta_{\mbox{\scriptsize\sc
ism}}) \approx (252^\circ, \, -17^\circ)$.  The values were chosen to
be consistent with~\cite{NFJ}; see references therein for citations
for these values.  The dipole anisotropy due to the Compton-Getting
effect is shown in Figs.~\ref{fig:2d_anisotropy_corrected}~(b) and
(c).  As mentioned in Section~\ref{sec:celestial}, the filter applied
to the data projects the dipole onto the equatorial plane, so the center of the
observed dipole has no declination component, and the effective
velocity is the equatorial projection of the velocity, which is
$v_{\mbox{\scriptsize\sc lsr}} \cos \delta_{\mbox{\scriptsize\sc lsr}}
= 17$~km/s for motion with respect to the local standard of rest, and
$v_{\mbox{\scriptsize\sc ism}} \cos \delta_{\mbox{\scriptsize\sc ism}}
= 21$~km/s for motion with respect to the interstellar medium.  The
cone parameters before and after subtracting the equatorial projection
of the Compton-Getting anisotropy from the observed anisotropy are
summarized in Table~\ref{tab:cone_params_cg_correct}.

We also examined the effect of the Compton-Getting anisotropy on the
one dimensional anisotropy.  Since the track- and zenith-type plots
are very similar, we focus here on just the zenith-type plot.
Starting with the two dimensional Compton-Getting anisotropy map
(Figs.~\ref{fig:2d_anisotropy_corrected}~(b) and (c)) and folding in
the effect of the overburden (Fig.~\ref{fig:overburden_norm}), the one
dimensional anisotropy due to the Compton-Getting effect alone is
described well by a first harmonic function with amplitude and phase
$(1.66 \times 10^{-4}, 274^\circ)$ for cosmic ray bulk motion with the
local standard of rest, and $(2.00 \times 10^{-4}, 256^\circ)$ for
bulk motion with the interstellar matter.  These two anisotropies are
indicated by the phasors $\vec{V}_{\mbox{\scriptsize\sc lsr}}$ and
$\vec{V}_{\mbox{\scriptsize\sc ism}}$ in
Fig.~\ref{fig:phasor_demo_cg_v3}~(b) and (c).  The amplitude and phase
of the observed anisotropy without any subtraction are indicated in
the figure as $\vec{D}$.  The contribution of the Compton-Getting
anisotropy is removed by subtracting $\vec{V}_{\mbox{\scriptsize\sc
lsr,ism}}$ from $\vec{D}$: the result is also shown in the figure.
The anisotropy is enhanced, and the direction of anisotropy rotates
toward $90^\circ$.  The effect of atmospheric subtraction is also
shown.  The amplitude and phase of the anisotropy before and after
subtraction are summarized in Table~\ref{tab:amp_phase_corrected}.
Also shown in the table are the anisotropies after subtracting both
effects.

\begin{table}
\begin{tabular}{|c|c|c|} \hline\hline
{\sc subtraction} & {\sc amplitude} & {\sc phase (deg)} \\
\hline\hline
{\sc none} & $5.7 \times 10^{-4}$ & $35^\circ$ \\ \hline\hline
{\sc cg, lsr} & $6.7 \times 10^{-4}$ & $48^\circ$ \\ \hline
{\sc cg, ism} & $7.3 \times 10^{-4}$ & $45^\circ$ \\ \hline
{\sc atmospheric} & $5.3 \times 10^{-4}$ & $40^\circ$ \\ \hline\hline
{\sc cg, lsr + atmospheric} & $6.5 \times 10^{-4}$ & $52^\circ$ \\ \hline
{\sc cg, ism + atmospheric} & $7.0 \times 10^{-4}$ & $50^\circ$ \\
\hline\hline
\end{tabular}
\caption{\label{tab:amp_phase_corrected} Amplitude and phase of the
  zenith-type one dimensional anisotropy before and after
  subtractions.  {\sc cg, lsr} refers to the Compton-Getting effect
  assuming cosmic ray rest frame moving with the local standard of
  rest, while {\sc cg, ism} refers to the case where the cosmic ray is
  assumed to move with the local interstellar matter.}
\end{table}

\section{The Clustering Algorithm}
\label{app:clustering_algorithm}

The clustering algorithm is applied to a histogram $N_{i,j}$, where
$N_{i,j}$ is the exposure-corrected number of events in bin ($i,j$),
$i = $ right ascension bin index, $j = $ declination bin index.  For
each ($i,j$), the quantity $\chi$ is calculated over a variable sized
cone centered on ($i,j$), where:

\begin{equation}
\chi = \frac{N_{obs} - N_{exp}}{\sqrt{N_{exp}}}
\end{equation}

\noindent $N_{obs}$ is the observed number of events in the cone, and
$N_{exp}$ is the expected number in the absence of anisotropy.  The
cone size (half opening angle) that extremizes $\chi$ is sought; the
excess/deficit is assumed significant if $\left| \chi \right| > 4$.
If the center of one cone falls within another cone, the cone with
small $\left| \chi \right|$ is rejected.

The statistical error of the reconstructed cone parameters was
estimated using an ensemble experiment technique in which an input sky
map with anisotropy cones with parameters given in
Table~\ref{tab:cone_params} was used to generate a large number of
output maps with random statistical fluctuations.  The clustering
algorithm was applied to each generated map, and the distribution of
the reconstructed cone parameters was examined.  The RMS of these
distributions were taken as the statistical error.


\bibliography{cr_anisotropy_paper_v26_2col}


\end{document}